\documentclass[aps,prb,twocolumn,groupedaddress,showpacs,floatfix,altaffilletter,longbibliography,includegraphics]{revtex4-2}
\usepackage{graphicx}
\usepackage{grffile}
\usepackage{amsmath}
\usepackage{amssymb}
\usepackage{bm}
\usepackage{color}
\usepackage[dvipsnames]{xcolor}
\usepackage{amsmath,amssymb,amsfonts}
\usepackage{epsfig}
\usepackage{times}
\usepackage[colorlinks,bookmarks=false,citecolor=blue,linkcolor=red,urlcolor=blue]{hyperref}
\usepackage{gensymb}
\usepackage[toc,page]{appendix}
\usepackage{float}
\usepackage[english]{babel}
\usepackage[T1]{fontenc}
\usepackage{comment}

\counterwithout{table}{section}

\newcommand{\figref}[1]{Fig.~\ref{#1}}

\newcommand{\secref}[1]{Sec.~(\ref{#1})}

\newcommand{\fref}[1]{Fig.~\ref{#1}}
\newcommand{\eref}[1]{Eq.~(\ref{#1})}

\DeclareUnicodeCharacter{202F}{\,} 
\DeclareUnicodeCharacter{202F}{ }  

\begin{document}

\title{Pseudogap, Fermi liquid, Van Hove singularity and maxima of the compressibility and of the Knight shift as a function of doping in the two-dimensional Hubbard model 
}

\author{ Y.M. Vilk}
\affiliation{Retired}
\author{A.-M.S.~Tremblay$^{1}$}
\affiliation{$^1$D\'epartement de physique \& Institut quantique \& RQMP\\
Universit\'e de Sherbrooke, Sherbrooke, Qu\'ebec, J1K 2R1 Canada}
\date{\today}

\begin{abstract} 
Qualitative changes in thermodynamic and single-particle properties characterize the transition between the pseudogapped electronic liquid and the Fermi liquid. Recent cold-atom experiments on a simulator of the Hubbard model with nearest-neighbor hoppings \cite{kendrick2025pseudogap} showed that the isothermal compressibility $\kappa(\delta)$ has a maximum as a function of doping $\delta$. Here we use the two-particle self-consistent plus (TPSC+) approach to explain these experiments and connect the maximum in $\kappa(\delta)$ to the transformation of the single-particle spectrum from the pseudogapped to the metallic regime. This elucidates the nature of the pseudogap (PG). Specifically, the maximum in $\kappa(\delta)$ practically coincides with the doping at which the precursor of the lower $(\pi,\pi)$ spin density wave (SDW) band at the antinodal point crosses the zero-frequency $\omega=0$. The Knight shift, $\chi_{sp}(0,0)(\delta)$, as a function of doping, should also have a maximum. In addition, we demonstrate that TPSC+ correctly predicts a maximum in the temperature dependence of the Knight shift, $\chi_{sp}(0,0)(T)$, consistent with recent ultracold atom experiments \cite{chalopin2026observation}.The maxima in both quantities should exist, at sufficiently low temperatures ($T$), in both the intermediate $U \approx U_{Mott}$ and weak $U < U_{Mott}$ interaction limits. In both limits, the mechanism is critical thermal SDW fluctuations. At the antinodal pseudogap, the correlation length at $\delta_{max}(T)$ can be small, controlled not by static but by dynamic critical thermal fluctuations. We also find that the SDW fluctuations are incommensurate at $\delta=\delta_{max}$. We predict that, at low $T$, the multiple peaks in the spin susceptibility in the incommensurate case lead to more than two SDW precursor peaks in the spectral function and density of states. By allowing access to parameter regimes relevant to cuprates—including further-neighbor hopping ($t', t''$) and low temperatures, our work provides a high-impact tool for further studies by the broader community.
\end{abstract}

\maketitle

\section{Introduction}
\label{sec:intro}
The decrease in the number of states at the Fermi level as temperature drops has become known as the pseudogap phenomenon.
The presence of pseudogap (PG) behavior has been confirmed in many quasi-two-dimensional materials, including all cuprate high $T_c$ superconductors (for reviews see \cite{Vishik2018,clement2025unraveling,Proust_Taillefer_2019}). 
There have been many proposals for the origin of the pseudogap in materials, as this small sample of references demonstrates 
\cite{kampf1990pseudogaps,Vilk1996, Vilk1997,varma1997non,norman1998phenomenology,franz1998phase,chen1998pairing,schmalian_microscopic_1999,senechal2004hot,kyung2006pseudogap,  Ferrero_Cornaglia_De_Leo_Parcollet_Kotliar_Georges_2009, LiebschTong:2009,Sordi_Haule_Tremblay_2010,  Gull_Ferrero_Parcollet_Georges_Millis_2010, sordi2012pseudogap, Gull_Millis_2013,  Sordi_Semon_Haule_tremblay_2013, Efetov_Meier_Pepin_2013, Pepin_de_Carvalho_Kloss_Montiel_2014, atkinson2015charge,Chowdhury_Sachdev, Fratino_Semon_Sordi_Tremblay_2016, Wu_Scheurer_Chatterjee_Sachdev_Georges_Ferrero_2018, Varma2019, wu2021interplay, Fratino_Bag_Camjayi_Civelli_Rozenberg_2021, gauvin2022disorder, Walsh_Charlebois_Sémon_Sordi_Tremblay_2022, wang2023phase, ye_location_2023, dai2020modeling,Ye_Chubukov_2023,sakai2023nonperturbative,kokkinis2025pseudogap,Wu_Scheurer_Chatterjee_Sachdev_Georges_Ferrero_2018,Bonetti_Christos_Nikolaenko_Patel_Sachdev_2025,Sachdev2020ancillaQbits,Meixner_Menke_Klett_Heinzelmann_Andergassen_Hansmann_Schafer_2024,lee2006}.
Spin-fluctuation-mediated theories of the pseudogap in the Hubbard model, connected to our approach, are compared in the discussion section of this paper, as is their possible relevance to cuprate physics.

We focus on the special case of the pseudogap appearing in the {\it nearest-neighbor one-band Hubbard model}.
%
%
As a starting point, we define the pseudogap as a momentum-dependent loss of spectral weight at the Fermi surface.

The recent availability of exact results from different methods makes it timely to aim for physical understanding and for a solution to all aspects of this problem, at least for this emblematic model of strongly correlated electrons~\cite{qin2022hubbard}.
In particular, a recent several-fold reduction in experimentally achievable temperatures \cite{xu2025neutral} enables detailed studies of the pseudogap phenomenon with cold atoms.
There should be agreement between analog cold-atom simulations of the Hubbard model~\cite{chalopin2026observation,kendrick2025pseudogap} and statistically exact diagrammatic (DiagMC) \cite{rossi2018directsamplingselfenergyconnected, Simkovic_Kozik_2019, Simkovic_Rossi_Georges_Ferrero_2024,vsimkovic2020extended,Schaefer2021} and determinantal (DQMC) quantum Monte Carlo \cite{Moukouri2000,schafer2015fate,Schaefer2021}. 
Other methods, such as infinite projected entangled pair states and minimally entangled typical thermal states, also provide reliable results~\cite{Sinha_Wietek_2025}.
Note that similar ranges of temperatures and interactions have become achievable with both cold atoms and Monte Carlo methods, another argument for the timeliness of our work. 
Accurate ground state studies~\cite{xu2022stripes} also provide guidance to investigate whether the pseudogap is a precursor of an ordered state or corresponds to a new ground state of matter, an important question. 

One of the most interesting questions related to the pseudogap phenomenon is the transition between the pseudogapped electronic liquid and the Fermi liquid. 
Several thermodynamic, one-particle, and two-particle physical properties are expected to change their behavior qualitatively at the boundary between these two distinct electronic regimes. 
In particular, recent cold-atom experiments \cite{kendrick2025pseudogap} showed that the isothermal compressibility has a maximum as a function of doping and connected this feature to the transition between the pseudogapped and Fermi-liquid states. 

The prediction of a maximum in the compressibility, $\kappa(\delta)$, in this transition from the pseudogapped to the correlated metallic states of the two-dimensional Hubbard model was first made in Ref.~\cite{sordi2012pseudogap} using cellular dynamical mean-field theory (CDMFT).
%
%
In addition to predicting the existence of the maximum in $\kappa(\delta)$, that work introduced the conceptual framework of the Widom line—originally developed in the context of fluids—to a completely different state of matter: the electronic fluid. 
The CDMFT approach is particularly effective in the strong-interaction limit.
A maximum in the isothermal compressibility as a function of doping for the near-neighbor Hubbard model was also recently observed theoretically in Ref.~\cite{Sinha_Wietek_2025}.

Here we use the two-particle self-consistent plus (TPSC+) approach \cite{Schaefer2021,gauvin2023improved,vilk2024antiferromagnetic}, to achieve the following:
\begin{itemize}
    \item 
a) We explain these recent cold atom experiments and directly connect the maximum in $\kappa(\delta)$ to the transformation of the single-particle spectrum from the pseudogapped to the correlated metallic regime.
Specifically, we show that the maximum in $\kappa(\delta)$, practically, coincides with the crossing of the precursor of the lower spin density wave (SDW) band in the spectral function $A(\mathbf{k}_{AN}, \omega)$ at the antinodal point from negative to positive energies. This, in turn, moves the Van Hove singularity in the density of states (DOS) from the occupied to the unoccupied side. The Van Hove singularity is very sensitive to the formation of a pseudogap~\cite{Wu_Scheurer_Ferrero_Georges_2020, vilk2023criteria}.
We illustrate possible finite-size effects on various observables in the cold-atom experiments.
\item 
b) We predict that the uniform magnetic susceptibility as a function of doping at fixed temperature $\chi_{sp}(0,0)(\delta)$, measured in the Knight shift, should also have a maximum for the same conditions. Numerically, the latter maximum may be easier to observe because it occurs at somewhat higher temperatures and does not require numerical differentiation. In addition, we demonstrate that TPSC+ correctly predicts a maximum in the temperature dependence of the Knight shift, $\chi_{sp}(0,0)(T)$, consistent with recent ultracold atom experiments \cite{chalopin2026observation}.   
A maximum of the Knight shift as a function of doping was also previously predicted in CDMFT~\cite{Sordi_Haule_Tremblay_2011}.
\item 
c) We predict that maxima in both $\kappa(\delta)$ and $\chi_{sp}(0,0)(\delta)$ should exist, at sufficiently low temperatures, in both intermediate $U \approx U_{Mott}$ and weak $U < U_{Mott}$ interaction limits. 
\item 
d)
In both limits, the mechanism that leads to the antinodal pseudogap, its disappearance with doping and concomitant maximum in $\kappa(\delta)$ and $\chi_{sp}(0,0)(\delta)$ is essentially the same, namely critical thermally excited {\it dynamical} SDW fluctuations (see below) with a correlation length of about a few lattice spacings, $\xi=4-5.5$, at $\delta_{max}$. 
\item 
e) The SDW fluctuations are already incommensurate at $\delta=\delta_{max}$. As the temperature decreases, $\delta_{max}$ moves toward the quantum critical SDW point, and the distance between the four incommensurate peaks in the spin susceptibility $\chi_{sp}(\mathbf{k},\omega) $ becomes larger. We predict that the multiple peaks in the spin susceptibility in the incommensurate case lead, at sufficiently low temperatures, to more than two precursor peaks in the spectral function and in the density of states.

%
\item 
g) The increase in the magnitude of the maximum of the compressibility at low temperatures implies that the charge susceptibility becomes large as well. 
This opens the possibility for a precursor of intertwined spin and charge density waves (CDW)~\cite{Meixner_Menke_Klett_Heinzelmann_Andergassen_Hansmann_Schafer_2024}, also known as stripes~\cite{Tranquada_2020,Fradkin_Kivelson_Tranquada_2015}, at a finite temperature. 
Stripes have been found in some ranges of parameters using finite-size ground state methods even in the Hubbard model~\cite{xu2022stripes}. 
However, finite temperature methods in the thermodynamic limit currently show no clear precursor of diverging charge susceptibility~\cite{qin2022hubbard,Simkovic_2021,xiao2023temperature,Simkovic_Rossi_Georges_Ferrero_2024,Sinha_Wietek_2025}. 
Our analysis indicates that, driven by critical SDW fluctuations, the precursor of CDW or of stripes at finite temperature may exist at very low temperatures deep in the pseudogap regime.
%
%
Comparisons with Ref.~\cite{xu2022stripes} suggest that the ground state is striped~\cite{Zheng_Chung_Corboz_Ehlers_Qin_Noack_Shi_White_Zhang_Chan_2017,Huang_Mendl_Liu_Johnston_Jiang_Moritz_Devereaux_2017,Mushkaev_Petocchi_Hoshino_Werner_2025,Devereaux_Kivelson_2025}, in which case our SDW mechanism for the pseudogap should crossover to stripe fluctuations as temperature is decreased deep in the pseudogap regime.
In that case, the question of the existence of a pseudogap ground state without long-range order would be answered negatively for that model. 
 \end{itemize}

 The paper is organized as follows. 
 The discussion of the method in \secref{sec:Method} begins with a motivation for the TPSC+ approach and a reminder of the criteria for the appearance of a pseudogap at different momentum points. 
 The section ends with  
 the model and a review of the TPSC and TPSC+ equations. 
 Our results for the compressibility in \secref{sec:Compressibility} begin in \secref{sec:Comparison_k_with_exp} with  
 comparisons with cold atom experiments based on calculations with a comparable system size.
 Finite-size effects on the maximum in $\kappa(\delta)$ in \secref{sec:Finite_size_effects},  have some relevance to cold atom experiments.
 Real-space correlation functions are also compared with cold-atom experiments in \secref{sec:Comparison_Cr_with_exp}.
 We end the discussion of the compressibility and its maximum as a function of doping in \secref{sec:TDL}, with results in the thermodynamic limit. 
 Results for the maximum of the uniform spin susceptibility as a function of doping in the thermodynamic limit are in \secref{sec:TDL_chisp00}. 
 In \secref{sec:N_om_and_A_k_om_evolution}, we discuss the evolution of single-particle properties with doping and show how they are related to the maxima in $\kappa(\delta)$ and $\chi_{sp}(0,0)(\delta)$ for both commensurate and incommensurate spin fluctuations.
 The section ends in \secref{sec:EDC_curves}, where we show the evolution of the spectral function $A_{\mathbf{k}}(\omega)$ as the momentum crosses the non-interacting Fermi surface (EDC curves). 
 In \secref{sec:dn_over_dm_vs_chich00}, we compare the compressibility $\kappa=\partial n/\partial \mu$ calculated using the second-level self-energy in TPSC+ with the uniform zero-frequency charge susceptibility $\chi_{ch}(0,0)$ at the first level of the TPSC+ approach. 
 In an exact theory, they should be identical, but here they differ.
 The second level of approximation should be better, so that is what we show in the main part of the paper.
 The disagreement allows us to argue for the possibility of charge density wave behavior deep in the pseudogap regime driven by SDW fluctuations. 
 The results for the dependence of the spin correlation length on doping in both the weak- and intermediate-interaction limits are shown in~\secref{sec:xi_vs_delta_two_U}, allowing a discussion of the similarities and differences between both cases.
 We end in \secref{sec:Discussion} with a discussion and comparison with results obtained with other methods and what they tell us about possible ground-state quantum critical points that lead to the observed behavior of observables discussed in this paper. 
 There, we also discuss the link to other spin-fluctuation-mediated theories of the pseudogap.
 We conclude in \secref{sec:conclusion}.

\section{Method}
\label{sec:Method}

We begin by explaining why the method that we use is appropriate for this problem and what it has already told us about the criteria for the appearance of a pseudogap at different momentum points. 
In the last subsection, we specify parameters of the model and the equations that are solved.

 \subsection{Motivation for using TPSC+ approach and criteria for the existence of the pseudogap at different momentum points}
\label{sec:TPSCpl_Criterion} 
  
 The TPSC+ method~\cite{Schaefer2021,gauvin2023improved,vilk2024antiferromagnetic} used in this paper is based on a generalisation of TPSC that was inspired by weak-interaction ideas.
 This extension of TPSC has been shown to work reasonably well up to the intermediate interaction regime ($U < 8$). Specifically, for $U < 3-3.5$ there is good quantitative agreement. For larger $U$, the TPSC+ is in qualitative agreement with DiagMC and Cold atom experiments data, but it underestimates pseudogap effects: qualitatively similar results appear in the TPSC+ at lower temperature for the same doping or lower doping at the same temperature. ~\cite{gauvin2023improved,vilk2024antiferromagnetic,vilk2025_crossovers}. 
 Some of the advantages of the TPSC+ approach are:
\begin{enumerate}
    \item It allows studying both the weak- and intermediate-interaction strengths within a single framework, providing a unified perspective on the problem.
		\item It satisfies the Pauli principle and sum rules for charge and spin fluctuations (fluctuation–dissipation theorems).
	\item It satisfies the Mermin–Wagner theorem, which precludes a finite-temperature antiferromagnetic (AFM) phase transition. 
    Instead of a true phase transition, it predicts a crossover to the renormalized classical (RC) regime with an exponentially growing correlation length as temperature decreases.
    Because of the large correlation length that diverges at $T=0$, we also call this regime ''the critical thermal fluctuation'' regime.

\item Predicts a pseudogap (PG) in the renormalized classical regime and identifies two analytical criteria for its existence: one for regular ($v_F \neq 0$) points on the Fermi surface (FS) \cite{Vilk1996, Vilk1997} and one for the antinodal (Van Hove, $v_F \sim 0$) point \cite{vilk2023criteria,vilk2025_crossovers}. 
\end{enumerate}

The criteria for the appearance of a pseudogap are important for the subsequent discussion. 
They were obtained analytically by considering thermally excited classical fluctuations in the RC regime. 
Even though all the results that we present are {\it based on numerical solution of the full set of equations of the next subsection}, physical insight is gained by the analytical solutions obtained in the RC regime with the Ornstein-Zernicke form of the spin fluctuations~\cite{Millis_Monien_Pines_1990}
\begin{equation}
    \chi(\mathbf{q},\omega)=\frac{A}{\xi^{-2}+(\mathbf{q}-\mathbf{Q})^2-i\omega/\gamma}.
\end{equation}
Here, $A$ is the amplitude of the susceptibility, $\xi$ is the spin correlation length, and $\gamma$ is the diffusion coefficient, or Landau damping, while $\mathbf{Q}$ is the wave vector where the susceptibility is maximum. 
In the incommensurate case, there are several maxima.

In natural units, $k_B=1$, $\hbar=1$, $a=1$, the criteria are:

\begin{enumerate}

\item
For a regular point on the FS ($v_F \gg \gamma/ \xi $) the criterion for the PG is~\cite{Vilk1996,Vilk1997}:
\begin{equation} \label{eq:Criteria_R}
\xi> \xi_{th\_db}=\frac{v_F}{\pi T}
\end{equation}
Here $\xi_{th\_db}$ is the thermal de Broglie wavelength of an electron. 
In practice, the PG appears when the ratio $\xi / \xi_{th\_db} \approx 1.5$. 
The above criterion is due to {\it static} ($\omega=0$) thermal fluctuations and becomes more favorable as the Fermi velocity decreases from the nodal point to the antinodal point.
\item
For the antinodal point on the FS and its vicinity ($v_F < \gamma/\xi$), dynamical effects, namely the frequency dependence of the thermal spin fluctuations, become important. 
In this regime, the criterion for the PG \eref{eq:Criteria_R} is inapplicable, and instead the criterion for the PG is~\cite{vilk2023criteria}:
\begin{equation} \label{eq:Criteria_AN}
\xi> \xi_{Dyn\_th}=\sqrt{\frac{\gamma}{\pi T}}.
\end{equation}
Here $\gamma$ is Landau damping, a dynamical characteristic of the spin fluctuations.  
\end{enumerate}

  The dynamical thermal length $\xi_{Dyn\_th}$ is significantly shorter than the thermal de Broglie wavelength $\xi_{th\_db}$, and thus the conditions for the appearance of the PG at the antinodal (Van Hove) point are significantly more favorable than at the nodal point, a theme we will encounter throughout this paper.

  The inequality between these two characteristic lengths increases with interaction strength $U$. 
  This happens because at larger $U$ the whole picture moves to larger temperatures ($T_{RC}$ increases) and Landau damping decreases as $\gamma \propto 1/T$ \cite{dare1996crossover,vilk2025_crossovers}.
  Taking into account this temperature dependence of $\gamma$, we have that the ratio between the two lengths increases as $T$ increases: 
  \begin{equation}
      \frac{\xi_{th\_db}}{\xi_{Dyn\_th}}=\frac{\frac{v_F}{\pi T}}{\sqrt{\frac{\gamma}{\pi T}}}\sim \sqrt{T}.
  \end{equation}
 
As temperature decreases and the correlation length grows, the pseudogap always appears first at the antinodal point at $T_{AN}$ and only significantly later at the nodal point at $T_{N}$. 
The inequality $T_{AN} > T_{N}$ becomes more pronounced as $U$ increases. 
For example, for $U=6.5$ and half-filling $\delta=0$, the PG at the antinodal point appears for $\xi \sim 1$ at $T_{AN}=0.37$, while the PG at the nodal point appears at $\xi \approx 8$ at $T_{N}=0.17$.
When doping increases at a given temperature, the pseudogap at the antinodal point is the last to disappear, as also found with diagrammatic quantum Monte Carlo~\cite{Simkovic_Rossi_Georges_Ferrero_2024}.

\subsection{Model and the two-particle self-consistent plus (TPSC+) approach:
TPSC+ Classical approach}
\label{sec:model}

\subsubsection{The Hubbard Model}
\label{sec:Hubbard Model}

We study the one-band nearest-neighbor Hubbard model on a two-dimensional square lattice, 
\begin{equation}
    H = \sum_{\mathbf{k},\sigma} \epsilon_{\mathbf{k}}\, 
    c^{\dagger}_{\mathbf{k}\sigma} c_{\mathbf{k}\sigma} 
    + U \sum_{i} n_{i\uparrow} n_{i\downarrow} 
    - \mu \sum_{i,\sigma} n_{i\sigma},
    \label{eq:hubbard}
\end{equation}
where $c^{\dagger}_{\mathbf{k}\sigma}$ ($c_{\mathbf{k}\sigma}$) creates 
(annihilates) an electron with spin $\sigma$ and wavevector $\mathbf{k}$, 
$n_{i\sigma}=c^{\dagger}_{i\sigma}c_{i\sigma}$ is the number operator for electrons with spin $\sigma$ 
at site $i$, $U$ is the on-site repulsive interaction, 
$\epsilon_{\mathbf{k}}$ is the bare electronic dispersion, 
and $\mu$ is the chemical potential.

Throughout this paper, we consider a square lattice with only 
nearest-neighbor hopping $t$. We set $t=1$, the lattice spacing $a = 1$, 
as well as $\hbar = 1$ and $k_B = 1$. In these units, 
the dispersion relation takes the form
\begin{equation}
    \epsilon_{\mathbf{k}} = -2(\cos k_x + \cos k_y).
    \label{eq:dispersion}
\end{equation}

\subsubsection{The TPSC+ Classical approach}
\label{sec:TPSC+}

In this subsection, we outline the TPSC+ classical approach and its application 
to the two-dimensional Hubbard model. 
The TPSC+ method extends the 
Two-Particle Self-Consistent (TPSC) theory, a non-perturbative approach that 
respects the Pauli principle, the Mermin-Wagner theorem, and sum rules for both spin and charge susceptibilities \cite{vilk1994theory, Vilk1997}.

The TPSC method has been extensively benchmarked against Monte Carlo simulations~\cite{TremblayMancini:2011, Schaefer2021} 
and provides a quantitatively accurate description in the weak to intermediate 
interaction regime over a broad range of doping levels and temperatures. 
Notably, TPSC was the first theoretical method to predict the opening of a 
 pseudogap arising from antiferromagnetic fluctuations {\it in the two-dimensional Hubbard model} within the renormalized classical regime, and it established the criterion for this phenomenon \cite{Vilk1996,Vilk1997}.
Methods that do not satisfy the Mermin-Wagner theorem cannot accomplish this.

While TPSC correctly captures the onset of the pseudogap in the renormalized 
classical regime, it fails deep within that regime. The TPSC+ approach \cite{Schaefer2021,gauvin2023improved,vilk2024antiferromagnetic} improves upon TPSC by incorporating feedback from single-particle properties into the two-particle quantities. This allows for the study of the pseudogap phenomenon at much lower temperatures and for larger correlation lengths.

Below, we briefly describe the TPSC+ formalism. The details can be found in Refs.~\cite{gauvin2023improved,vilk2024antiferromagnetic}.

The spin and charge susceptibilities take RPA-like forms, but with the bare 
interaction $U$ replaced by the renormalized spin ($U_{sp}$) and charge 
($U_{\text{ch}}$) interactions. Additionally, the bare bubble is replaced by 
a semi-dressed bubble, which will be defined shortly below:

\begin{align}
    \chi_{sp}(\mathbf{q},iq_n) = \frac{\chi^{(2)}(\mathbf{q},iq_n)}{1-\frac{U_{sp}}{2}\chi^{(2)}(\mathbf{q},iq_n)},
    \label{eq:chisp2}\\
    \chi_{ch}(\mathbf{q},iq_n) = \frac{\chi^{(2)}(\mathbf{q},iq_n)}{1+\frac{U_{ch}}{2}\chi^{(2)}(\mathbf{q},iq_n)}.
    \label{eq:chich2}
\end{align}
Here $\mathbf{q}$ is the momentum and $q_n=2\pi n $ are Bosonic Matsubara frequencies.  The spin and charge irreducible vertices are computed in the same way as in the TPSC approach, namely through the self-consistency with the local sum rules which have the form~\cite{vilk1994theory}

\begin{align}
\frac{T}{N}\sum_{\mathbf{q},q_n} \chi_{sp}(\mathbf{q},iq_n) &= n-2 \langle n_{\uparrow}n_{\downarrow} \rangle, \label{eq:sumrule_sp}\\
\frac{T}{N}\sum_{\mathbf{q},q_n} \chi_{ch}(\mathbf{q},iq_n) &= n+2 \langle n_{\uparrow}n_{\downarrow} \rangle - n^2.  \label{eq:sumrule_ch}
\end{align}
and are special cases of the fluctuation-dissipation theorem (FDT).

 Note that to arrive at the expressions on the right one needs to use the Pauli principle:

\begin{equation}
 n_{\sigma}^2=n_{\sigma}
\label{eq:Pauli_principle} 
\end{equation}

To compute self-consistently the vertices $U_{sp}$ and $U_{ch}$ from the sum rules, a third equation is needed. In  TPSC and TPSC+, the following \emph{ansatz}~\cite{hedayati1989ground,vilk1994theory} for  $U_{sp}$ is used:

\begin{equation}
U_{sp} = U \frac{\langle n_{\uparrow}n_{\downarrow} \rangle}{\langle n_{\uparrow} \rangle \langle n_{\downarrow} \rangle}.  \label{eq:ansatz_tpsc}
\end{equation}
 
The renormalized vertex $U_{sp}$ is suppressed relative to the bare 
interaction due to the influence of the pair correlation function 
$g_{\uparrow \downarrow} (0) = \langle n_{\uparrow} n_{\downarrow} \rangle / 
\langle n_{\uparrow} \rangle \langle n_{\downarrow} \rangle$. The RPA 
approximation is recovered when $g_{\uparrow \downarrow} (0) = 1$, while the deviation of 
$g_{\uparrow \downarrow} (0)$ from unity accounts for Kanamori-Br\"uckner 
screening.

 The semi-dressed bubble $\chi^{(2)}(\mathbf{q}, i q_n)$ in the TPSC+ classical 
approach is calculated as
\begin{align}
    \chi^{(2)}(\mathbf{q},iq_n&) =  \nonumber \\
    -\frac{T}{N}&\sum_{\mathbf{k},ik_n}\left((\mathcal{G}_{\sigma}^{(2)}(\mathbf{k},ik_n)\mathcal{G}_{\sigma}^{(1)}(\mathbf{k}+\mathbf{q},ik_n+iq_n)\right.\nonumber \\
     &+\left. \mathcal{G}_{\sigma}^{(2)}(\mathbf{k},ik_n)\mathcal{G}_{\sigma}^{(1)}(\mathbf{k}-\mathbf{q},ik_n-iq_n) \right ),
    \label{eq:chi2}
\end{align}
where $\mathcal{G}_{\sigma}^{(1)}(\mathbf{k} - \mathbf{q}, i k_n - i q_n)$ 
is the non-interacting Green function, and 
$\mathcal{G}_{\sigma}^{(2)}(\mathbf{k}, i k_n)$ is the interacting Green 
function, calculated iteratively using the self-energy arising from the static classical 
spin fluctuations responsible for the pseudogap deep in the RC regime:

\begin{align}
    \Sigma_{cl}(\mathbf{k},ik_n) = \frac{T}{N} \tilde{g}_{fb} U U_{sp} \sum_{\mathbf{q}} \chi_{sp}(\mathbf{q},0) \mathcal{G}_\sigma^{(1)}(\mathbf{k}+\mathbf{q},ik_n).		
    \label{eq:selfEnergy_Cl}
\end{align} 

Here $\tilde{g}_{fb}$ is a numerical factor that we will discuss later and $k_n=(2n+1) \pi$ are Fermionic frequencies. The approach is analogous to the pairing approximation (GG0 theory) for the pair susceptibility originally developed by Kadanoff and 
Martin~\cite{Kadanoff_Martin_1961, chen2005bcs, Boyack_2018}. 

There are several advantages to this approach:

\begin{enumerate}
    \item While the non-interacting bubble diverges as $T \to 0 $ in the 
    half-filled case -thereby pushing $U_{sp}$ to zero in TPSC- the  
		semi-dressed bubble $\chi^{(2)}(\mathbf{q}_{max}, 0)$ does not 
    diverge. Instead, in the pseudogap regime, it becomes exponentially 
    close to $2 / U_{sp}$, and both quantities converge to a finite 
    value at zero temperature.
  \item The condition 
    $\chi^{(2)}(\mathbf{q}_{max}, 0) U_{sp} / 2 \approx 1$ 
    leads to a modified Stoner criterion for the gap value, where the 
    bare $U$ is replaced by the renormalized interaction $U_{sp}$.
    
\end{enumerate}

The motivation for using the TPSC+ classical approach is that it employs the 
minimal necessary feedback self-energy to achieve a reliable description of 
spin susceptibility deep in the renormalized classical regime. For simplicity, we will refer to TPSC+ Classical simply as TPSC+ in this paper.

 The full self-energy is computed in a second step, using the spin and charge 
susceptibilities obtained in the first step, which incorporate collective 
modes~\cite{Vilk1996, Vilk1997, Moukouri2000}:

\begin{align}
    \Sigma_\sigma^{(2)}(\mathbf{k},&ik_n) = Un_{-\sigma}\nonumber+\frac{T}{N} U\sum_{\mathbf{q},iq_n}\left [\tilde{g} U_{sp}\chi_{sp}(\mathbf{q},iq_n)\right.\nonumber \\
     &+\left.(1/2 - \tilde{g}) U_{ch}\chi_{ch}(\mathbf{q},iq_n)\right ] \mathcal{G}_\sigma^{(1)}(\mathbf{k}+\mathbf{q},ik_n+iq_n).		
    \label{eq:selfEnergy2}
\end{align} 

Contrary to the electron-phonon case, there is no Migdal theorem for 
electron-electron interactions; spin and charge vertices are renormalized.

In the original TPSC papers~\cite{Vilk1996, Vilk1997}, the value 
$\tilde{g} = 1/4$ was adopted. A subsequent study~\cite{Moukouri2000} proposed 
that $\tilde{g} = 3/8$ is a more appropriate choice. This discrepancy stems 
from the fact that using a constant vertex $U_{\text{sp}}$ breaks rotational 
invariance, leading to distinct formulations in the longitudinal and 
transverse spin channels. The value $\tilde{g} = 3/8$ is derived by averaging 
over both channels in an attempt to restore crossing symmetry (i.e., rotational 
invariance). By comparing to the Monte Carlo results,we found that for $U <3$ the combination $\tilde{g}_{fb} = 3/8$ and $\tilde{g} = 1/4$ work best, while for $U \geq 3$ the combination $\tilde{g}_{fb} = 1/4$ and $\tilde{g} = 3/8$ work best~\cite{vilk2024antiferromagnetic,vilk2025_crossovers}. Here, the factor $\tilde{g}_{fb}$ is used in the feedback self-energy \eref{eq:selfEnergy_Cl} and the factor $\tilde{g}$ used in the expression for the full self-energy \eref{eq:selfEnergy2}.

The program for the self-consistent solution of the above TPSC+ equations runs in less than 2 minutes on a laptop, thanks in large part to the use of fast Fourier transforms for momentum convolutions and sparse-IR for Matsubara-frequency convolutions~\cite{Shinaoka_2017, Shinaoka_sparse_2022, Li_2020}. The initial Python program was developed by Chloé Gauvin-Ndiaye and Camille Lahaie \cite{gauvin2023improved}. Further improvements were made in \cite{vilk2024antiferromagnetic}, where the convergence problem at low temperatures was resolved, and the capability to calculate very large correlation lengths ($\xi < 10^6$) was added. 

Finally, we stress that, given the high accuracy of our numerical calculations and the absence of noise in Matsubara data, all the results involving analytical continuation could be obtained using Padé approximants. 

\section{Compressibility results}
\label{sec:Compressibility}

A faithful comparison with experiments requires calculations made on comparable system sizes. The subsection that follows these comparisons studies finite system sizes to clarify these effects. Real-space correlations, also relevant for experiments, are presented in the next-to-last subsection. We end this section with results in the thermodynamic limit. 

\subsection{Comparison of TPSC+ results for the compressibility with cold atoms experiments}
\label{sec:Comparison_k_with_exp}

We start by noting a technical point. In the cold atom paper \cite{kendrick2025pseudogap}, the authors used a simplified version of the compressibility definition, $\kappa=\partial n/\partial \mu$. The thermodynamic definition is $\kappa=1/n^2\partial n/\partial \mu$, which is the one used in the paper on the Widom line \cite{sordi2012pseudogap}. The difference between these definitions affects the exact shape of the curve $\kappa(n)$ but does not qualitatively change the results. To simplify comparison with the experimental results in Ref.~\cite{kendrick2025pseudogap}, we used the definition from that paper, i.e., $k=\partial n/\partial \mu$.

\begin{figure}
    \centering
    \includegraphics[width=\columnwidth]{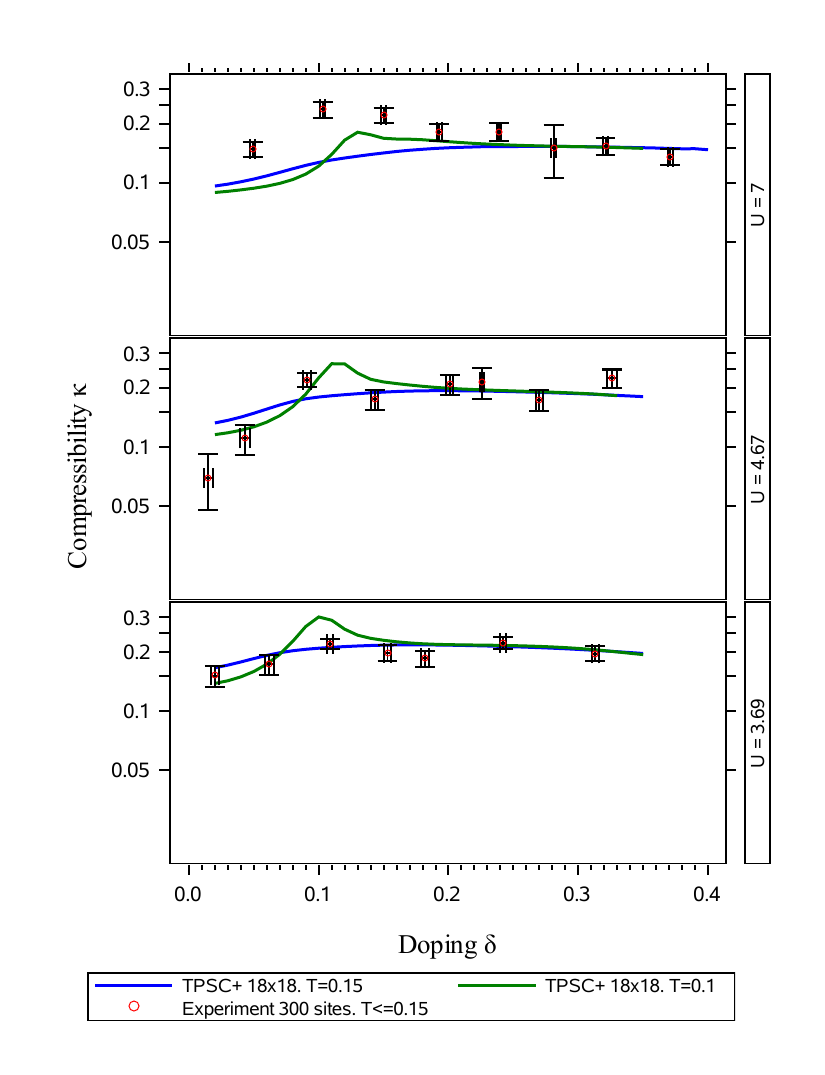} 
    \caption{ Comparison of the TPSC+ doping dependence of the compressibility $\kappa(\delta)$ with results from experiments with a cold-atom simulator of the Hubbard model. The experiments were conducted on a system of $300$ sites in a circular geometry, while TPSC+ calculations were performed on an $18\times18$ lattice (324 sites) with periodic boundary conditions. The three interaction strengths, $U=(3.69,4.67,7)$, correspond to weak-to-intermediate interactions. The temperature in the experiment is $T \lesssim 0.15$, and the TPSC+ calculations were performed for two temperatures: $T=0.1$ and $T=0.15$.  The TPSC+ theory predicts the existence of a well-defined maximum in the compressibility as a function of doping $\kappa(\delta)$ for $T=0.1$ for all considered $U$. By contrast, there is only a broad maximum in $\kappa(\delta)$ for $T=0.15$. In the weaker interaction case $U=3.69$, there is quantitative agreement with the experiment, while in the intermediate interaction case $U=7$, the theory somewhat underestimates $\kappa(\delta)$ around the maximum. This is consistent with other observations indicating that TPSC+ underestimates the antiferromagnetic correlations in the intermediate-interaction limit.}
    \label{fig:Compes_k_by_U_T_comp_exp}
\end{figure}

\fref{fig:Compes_k_by_U_T_comp_exp} shows a comparison of the TPSC+ doping dependence of the compressibility $\kappa(\delta)$ with experimental results from a cold-atom simulator of the Hubbard model \cite{kendrick2025pseudogap}. The experiments were conducted on a system of $300$ sites in a circular geometry, while the TPSC+ calculations were performed on an $18\times18$ lattice (324 sites) with periodic boundary conditions. The three interaction strengths, $U = (3.69, 4.67, 7)$, span the weak-to-intermediate interaction regimes. The experimental temperature was determined to lie within a range $T \lesssim 0.15$, depending on the method used for temperature calibration (see Fig. S5 in Ref.~\cite{kendrick2025pseudogap}). The uncertainty in temperature also depends on the interaction strength, but for simplicity, we compare the experimental data with TPSC+ results for two temperatures: $T = 0.1$ and $T = 0.15$. The conclusion is that in the weaker interaction case $U=3.69$, there is quantitative agreement with the experiment, while in the intermediate interaction case $U=7$, the theory somewhat underestimates $\kappa(\delta)$ around the maximum. This is consistent with other observations indicating that TPSC+ underestimates the antiferromagnetic correlations in this limit. A direct comparison of the spin correlation function with cold-atom experiments for $U = 7$ is discussed below in \secref{sec:Comparison_Cr_with_exp}.

The TPSC+ theory predicts the existence of a well-defined maximum in the compressibility as a function of doping, $\kappa(\delta)$, for $T = 0.1$ for all interaction strengths considered. By contrast, only a very broad maximum in $\kappa(\delta)$ remains at $T = 0.15$. We will discuss the origin of this behavior in \secref{sec:N_om_and_A_k_om_evolution}, where we connect the evolution of the single-particle spectra with doping to the emergence of the maximum in $\kappa(\delta)$.

\subsection{Finite system-size effects on the maximum in $\kappa(\delta)$ }
\label{sec:Finite_size_effects}

\begin{figure}
    \centering
    \includegraphics[width=\columnwidth]{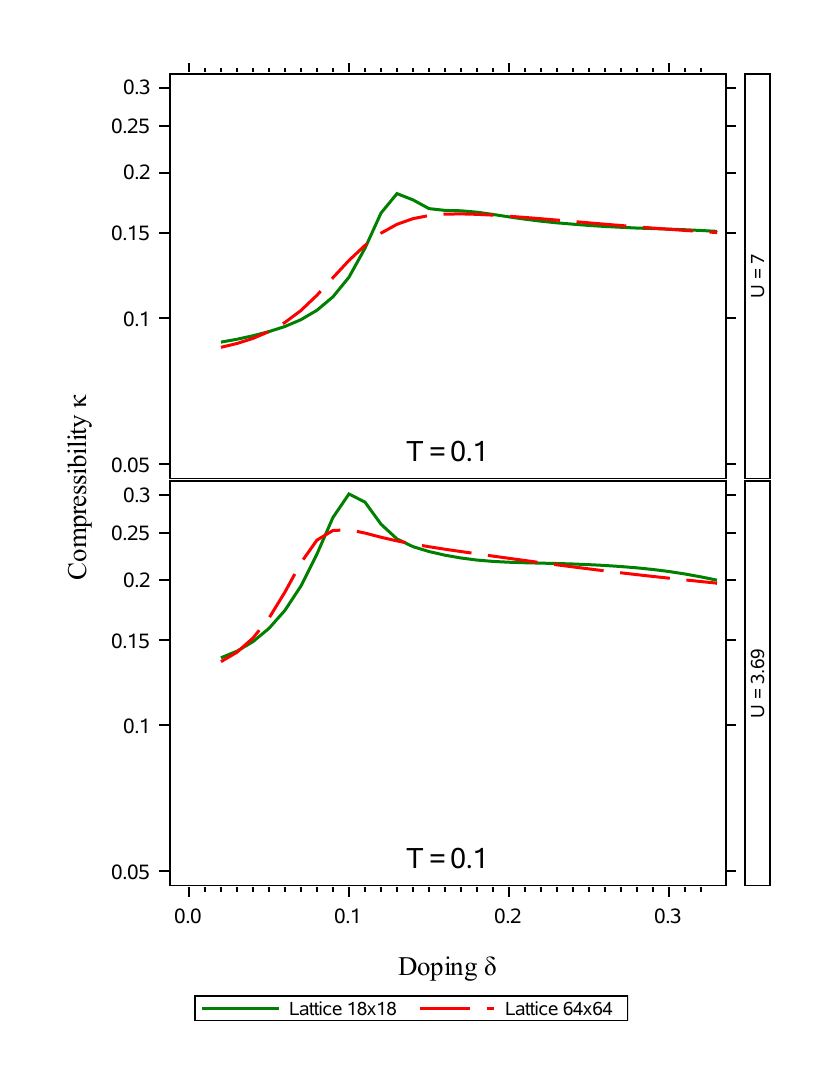} 
    \caption{Effect of the finite system size on the doping dependence of the compressibility $\kappa(\delta)$. The temperature is $T=0.1$, the two interaction strengths are $U=(3.69, 7)$ and the system sizes are $18\times18$ and $64\times64$. We can see that $\kappa(\delta)$ has a significantly sharper maximum in the $18\times18$ system than in the $64\times64$ system. The correlation length for doping close to the maximum $\delta\sim \delta_{max}$ is a few lattice spacings in this case. This affects the result for $18\times18$ but not for the $64\times64$ system. We double checked that increasing the linear system size $L$ above $L=64$ has virtually no effect on the maximum in the $\kappa(\delta)$. The analysis suggests that the finite size effect on the maximum in $\kappa(\delta)$ plays a significant role when about a quarter of the linear system size is smaller than the correlation length $1/4 L < \xi$.}
    \label{fig:Compes_k_by_U_size_effect}
\end{figure}

 In this section, we investigate the effect of finite system size on the doping dependence of the compressibility $\kappa(\delta)$ using the TPSC+ approach. The details are explained in the figure caption for \fref{fig:Compes_k_by_U_size_effect}. The main conclusion is that finite-size effects enhance the maximum in the doping dependence of the compressibility $\kappa(\delta)$ when about a quarter of the linear system size is smaller than the correlation length, $L/4 < \xi$. As we show below in \secref{sec:TDL}, the sharp maximum in $\kappa(\delta)$ does exist in the thermodynamic limit for the $U=(3.69, 7)$ but at a lower temperature, $T=1/14 \approx 0.0714$. Thus, the small system size effect is qualitatively similar to the effect of lower temperature in the thermodynamic limit. The above finite-size effect can be understood using the TPSC+ approach. There are two related reasons for that:

\begin{enumerate}
    \item In the expression for the self-energy in terms of the spin fluctuations \eref{eq:selfEnergy2}, the contribution of the AFM vector $\mathbf{Q}$ has relatively larger weight in the smaller system.
		
  \item In the expression for the density of states $N(\omega)=1/N \sum A(\mathbf{k}, \omega)$, the relative weight of the antinodal, Van Hove point $\mathbf{k}=(\pi,0)$, is larger in the smaller system.
\end{enumerate}      

\subsection{Comparison of the TPSC+ results for the real space correlation function with cold atoms experiments}
\label{sec:Comparison_Cr_with_exp} 

In this section, we compare TPSC+ results for the real-space spin correlation function at equal times with the results of experiments on a cold-atom Hubbard model simulator \cite{kendrick2025pseudogap}.

\begin{figure}
    \centering
    \includegraphics[width=\columnwidth]{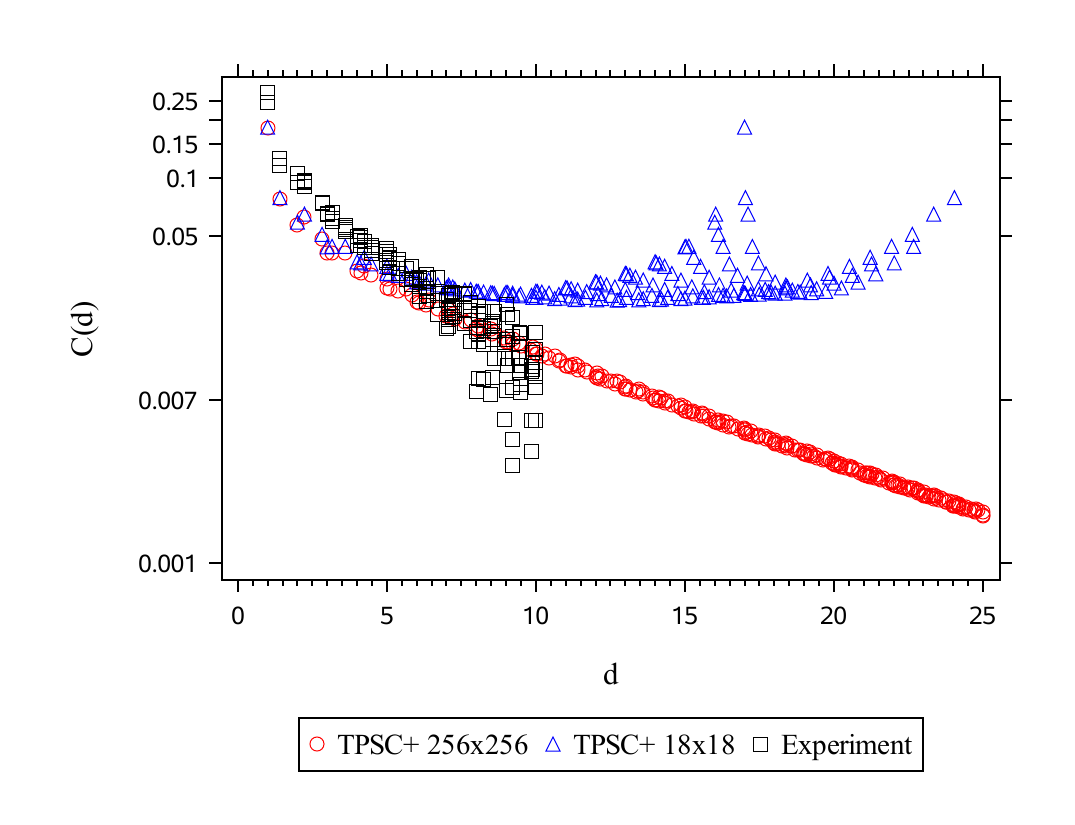} 
     \caption{ Comparison of the experimental equal-time spin correlation function $|C(d)|$ as a function of Euclidean distance $d$ with TPSC+ results for two lattices, $18 \times 18$ and $256 \times 256$. The parameters are $U=7$, $T=0.15$ and $\delta=0.04$. The first lattice has approximately the same number of sites as in the experiment but a different geometry (square with periodic boundary conditions versus circular, without such conditions). The second lattice is equivalent to the thermodynamic limit (TDL), having a TPSC+ correlation length of $\xi=9.74$. The finite-size effects are negligible in both small systems for short distances, $d \leq 6 \sim L/3$, but become progressively more important for larger distances. For $d \leq 6$, the TPSC+ results underestimate the spin correlation function by about $27\%$ to $33\%$ for $U=7$. This is consistent with our earlier observation that TPSC+ underestimates spin correlations in the intermediate interaction case. The fits to theoretical and experimental data are described in the main text.}
    \label{fig:Cr_func_U_7_delta_0_04_T_0_15}
\end{figure}

\fref{fig:Cr_func_U_7_delta_0_04_T_0_15} shows the comparison between TPSC+ and experiment for the absolute value of the spin correlation function $|C(\mathbf{d})|$ on a semi-logarithmic scale. The parameter values are $U \approx 7$, $T \approx 0.16$, and $\delta = 0.04$ for this figure in \cite{kendrick2025pseudogap}. There is some uncertainty in the values of $U$ and $T$ because experimentally, different calibration methods yield slightly different results.

The TPSC+ results are presented for two system sizes: the $18 \times 18$ system, which has 324 sites—close to the 300 sites used in the experiment-and the $256 \times 256$ system, which in this case provides an excellent approximation to the thermodynamic limit (TDL). The parameters for the TPSC+ calculations are $U=7$, $T=0.15$, and $\delta=0.04$. The correlation length estimated using TPSC+ in the TDL is $\xi=9.74$.

Before we discuss the comparison with experiment, let us first examine the finite-size effects on the results for $|C(d)|$. From the TPSC+ results in \fref{fig:Cr_func_U_7_delta_0_04_T_0_15}, one can see that size effects become significant in the smaller $18 \times 18$ system for distances $d \geq 6 \approx L/3$. In other words, only short-range correlations with $d < L/3$ can be faithfully reproduced by a finite system.

One indication of finite-size effects is the increasing spread of the $|C(d)|$ values for a given $d$ when $d > L/3$. In an infinite system, the asymptotic behavior of the spin correlation function $|C(\mathbf{d})|$ for $d \gg 1$ depends only on the absolute value of $\mathbf{d}$. However, in a finite system, the function $C(\mathbf{d})$ is not a unique function of the distance $d$ near the system boundary. This causes the increasing spread of $|C(d)|$ values for a given distance $d$, as observed in both experimental and theoretical small systems in \fref{fig:Cr_func_U_7_delta_0_04_T_0_15}. See~\footnotemark[1] for further details.

A comparison of the experimental data for $C(d)$ with $d \leq 6$ and the TPSC+ calculations shows that TPSC+ underestimates the spin correlations by about $27\%$ to $33\%$ for $U=7$. This is consistent with what we observed earlier in the intermediate-interaction case. Such an underestimation explains the quantitative disagreement between the TPSC+ data for $\kappa(\delta)$ and the experimental results in the intermediate-interaction regime (see \secref{sec:Comparison_k_with_exp}).

\footnotetext[1]{Due to differences in geometry and boundary conditions, finite-size effects manifest themselves somewhat differently in theoretical and experimental small systems. In particular, in a small theoretical system, $C(d)$ exhibits a minimum as a consequence of the periodic boundary conditions, whereas small experimental systems do not show such a minimum. However, in both cases, the spread of the $C(d)$ values for the same $d$ increases with distance, indicating significant finite-size effects for $d > L/3$.}

\subsection{Maximum of the compressibility as a function of doping in the thermodynamic limit }
\label{sec:TDL}   

\begin{figure}
    \centering
    \includegraphics[width=\columnwidth]{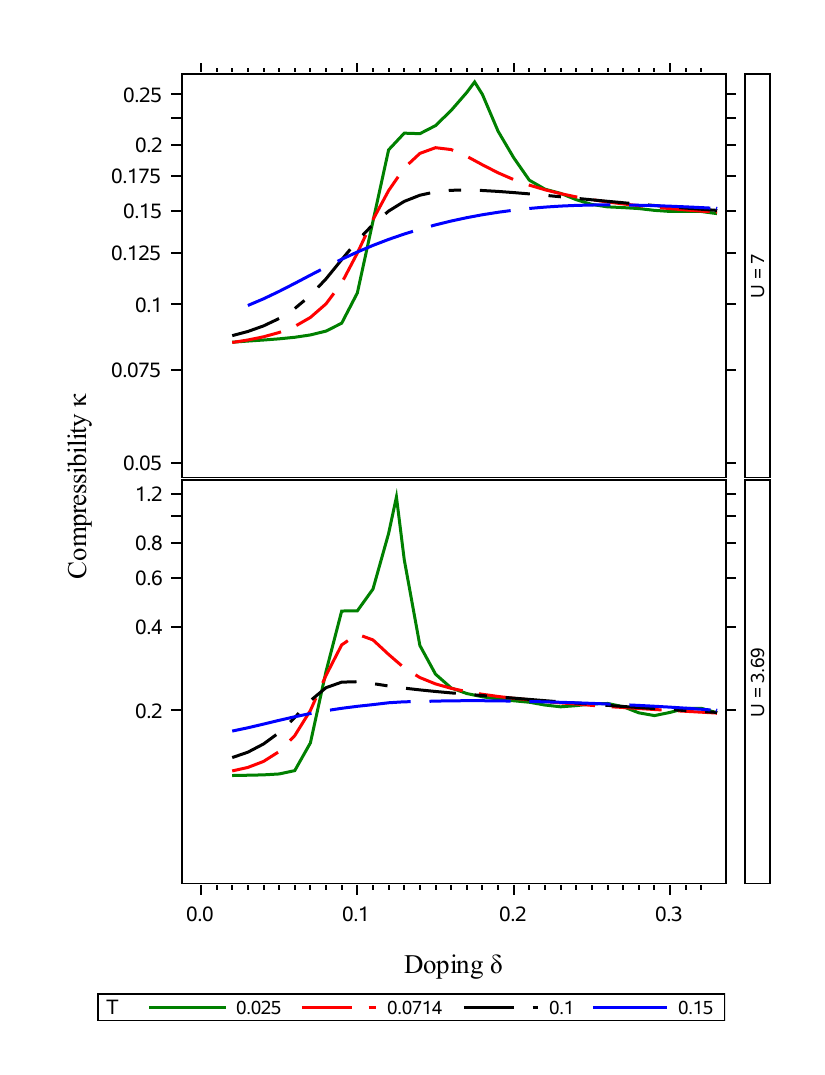} 
    \caption{Dependence of the compressibility $\kappa$ on doping for four different temperatures, $T = (0.15, 0.1, 0.0714,0.025)$, and two different interaction strengths, $U = (3.69, 7)$. The smaller interaction strength represents a weak-interaction case, $U = 3.69 <  U_{Mott}$, while the larger interaction strength, $U = 7 \approx  U_{Mott}$, corresponds to an intermediate-interaction case. The results are obtained in the thermodynamic limit. The doping-dependent isothermal compressibility $\kappa(\delta)$ shows a maximum for $T = 0.1 $. The curves for both interaction strengths $U$ are qualitatively similar, indicating that the same physics is at play. The maximum in $\kappa(\delta)$ signals the crossover from a pseudogapped electronic liquid to a correlated Fermi liquid at $\delta_{max}(T)$. The value of $\delta_{max}$ is significantly lower for smaller $U$ (for $T = 1/14 \approx 0.0714$, we have $\delta_{max} = 0.1$ for $U = 3.69$ while $\delta_{max} = 0.15$ for $U = 7$). }   
    \label{fig:Compes_k_by_U_T_inf_Lattice}
\end{figure}

\fref{fig:Compes_k_by_U_T_inf_Lattice} shows the dependence of the compressibility $\kappa$ on doping $\delta$ for four different temperatures, $T = (0.15, 0.1, 0.0714, 0.025)$, and two different interaction strengths, $U = (3.69, 7)$. The detailed results are described in the figure caption. The main conclusion is that a maximum in $\kappa(\delta)$ is predicted by TPSC+ in the thermodynamic limit for $T \leq 0.1$. The results are qualitatively similar to the Widom line paper \cite{sordi2012pseudogap}. However, TPSC+ suggests a different perspective: the effect is observed in both intermediate and weak interaction limits and is caused by SDW fluctuations. This suggests that Mott physics is not necessary for the effect. This is discussed further in Sec.~\ref{sec:Discussion}.

The $\kappa(\delta)$ results for the lowest considered temperature $T=0.025$ show additional structure near the maximum that is absent at higher temperatures. As doping increases, there is an initial rise in $\kappa(\delta)$ followed by a short plateau, which is then followed by a sharp maximum. As will be explained in \secref{sec:incommensurate_case}, the more complex structure of the maximum in $\kappa(\delta)$ is caused by two factors: the strongly incommensurate nature of the critical spin fluctuations and the existence of the pseudogap at both nodal $\mathbf{k}_N=(\pi/2, \pi/2)$ and antinodal points $\mathbf{k}_{AN}=(\pi,0)$. By contrast, the maximum in $\kappa(\delta)$ at higher temperatures is driven only by the pseudogap at the antinodal point.      

Most of the results in this section were obtained using a $64 \times 64$ lattice, but we verified a sample of data points using $256 \times 256$ and $512 \times 512$ lattices and found virtually no difference.

As illustrated in \secref{sec:Comparison_k_with_exp}, for smaller systems (e.g., $18 \times 18$), which are similar in size to those studied in cold atom experiments (about 300 sites), there is significant system-size dependence in the region of the curve $\kappa(\delta)$ near the maximum $\delta_{max}$. However, close to $\delta_{max}$, the correlation length is only several lattice spacings (for $T = 1/14 = 0.0714$, $\xi_{max} = 4$ for $U = 7$ and $\xi_{max} = 5.6$ for $U = 3.69$). Thus, lattices with system size $L \ge 64$ are sufficient to obtain results representative of the thermodynamic limit (TDL).

\section{Maximum of the Knight shift as a function of doping in the thermodynamic limit }
\label{sec:TDL_chisp00} 
 
\begin{figure}
    \centering
    \includegraphics[width=\columnwidth]{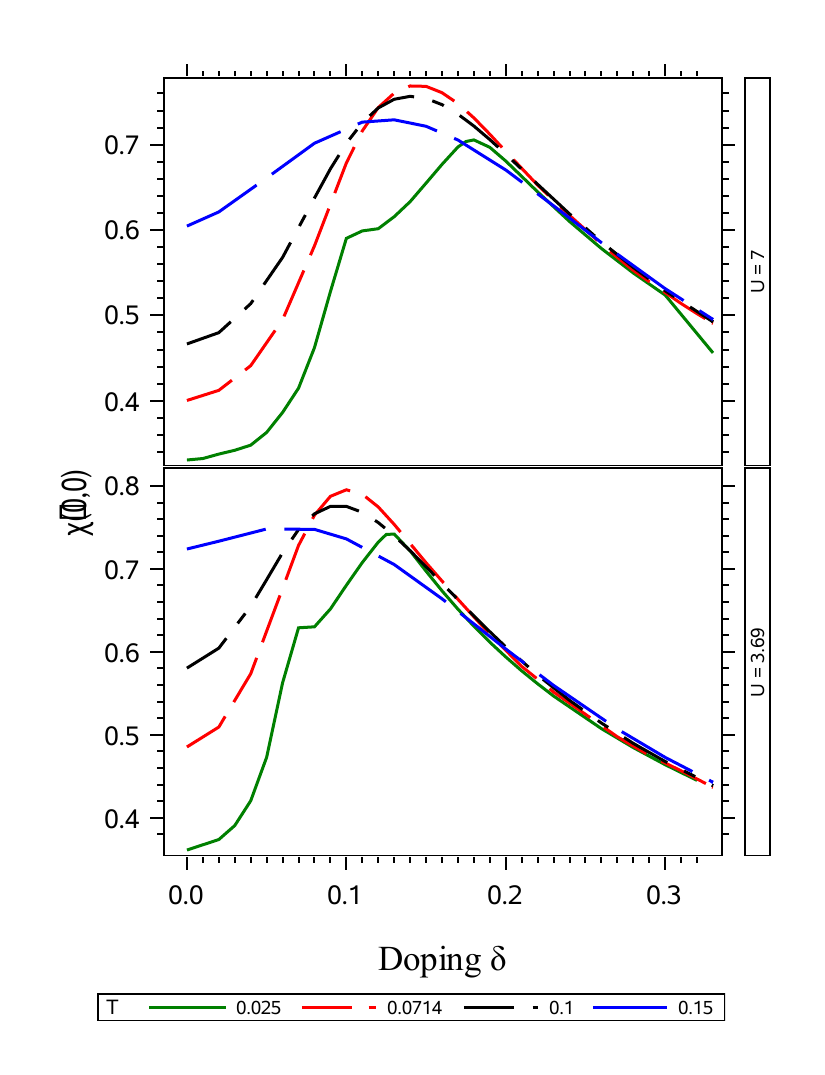} 
    \caption{Dependence of the uniform spin susceptibility $\chi_{sp}(0,0)$ on doping for four different temperatures, $T = (0.15, 0.1, 0.0714,0.025)$, and two different interaction strengths, $U = (3.69, 7)$. The smaller interaction strength represents a weak-interaction case, $U = 3.69 <  U_{Mott}$, while the larger interaction strength, $U = 7 \approx  U_{Mott}$, corresponds to an intermediate-interaction case. The results are obtained in the thermodynamic limit. The function $\chi_{sp}(0,0)(\delta)$ shows a maximum for all temperatures. The curves for both interaction strengths $U$ are qualitatively similar, indicating that the same physics is at play. The maximum in $\chi_{sp}(0,0)(\delta)$signals the transition from a pseudogapped electronic liquid to a correlated Fermi liquid and lies on the crossover line $\delta_{max}(T)$. The value of $\delta_{max}$ is significantly smaller for smaller $U$ (for $T = 1/14 \approx 0.0714$, $\delta_{max} = 0.1$ for $U = 3.69$ while $\delta_{max} = 0.15$ for $U = 7$).}   
    \label{fig:chisp00_vs_delta_by_U_T}
\end{figure}

\fref{fig:chisp00_vs_delta_by_U_T} shows the dependence of the uniform spin susceptibility $\chi_{sp}(0,0)$ (Knight shift) on doping for four different temperatures, $T = (0.15, 0.1, 0.0714, 0.025)$, and two interaction strengths, $U = (3.69, 7)$. The smaller interaction strength, $U = 3.69 < U_{Mott}$, corresponds to a weak-interaction case, while the larger value, $U = 7 \approx U_{Mott}$, represents an intermediate-interaction case. All results are obtained in the thermodynamic limit. The function $\chi_{sp}(0,0)(\delta)$ exhibits a maximum at all temperatures shown.

For low temperatures $T \leq 0.1$, the maximum in $\chi_{sp}(0,0)(\delta)$ occurs at essentially the same doping values as the maximum of the compressibility $\kappa(\delta)$. However, at the higher temperature $T = 0.15$, $\chi_{sp}(0,0)(\delta)$ still displays a well-defined maximum, while $\kappa(\delta)$ does not. The origin of this difference is discussed in \secref{sec:N_om_and_A_k_om_evolution}. We verified that a maximum in $\chi_{sp}(0,0)(\delta)$ appears at some doping value $\delta_{max}(T)$ when a pseudogap exists on part of the  Fermi surface at half-filling ($\delta=0$). As the temperature decreases, $\delta_{max}(T)$ moves away from half-filling ($\delta = 0$) and shifts toward the quantum critical point of the spin density wave (SDW) transition. That quantum critical point is very close to that found in Ref.~\cite{xu2022stripes}.

\section{Maximum of the Knight shift as a function of temperature and comparison with ultracold atom experiments and virtually exact numerical methods}
\label{sec:chisp00_vs_T} 

One of the first indications of the pseudogap in high-$T_c$ superconductors was the observation of a maximum in the temperature dependence of the uniform magnetic susceptibility (Knight shift), $\chi_{sp}(0,0)$ \cite{johnston1989magnetic,Alloul_Ohno_Mendels_1989}. In this section, we show that TPSC+ correctly predicts the existence of such a maximum in the Hubbard model and compare our results for the uniform magnetic susceptibility with ultracold atom experiments, determinant quantum Monte Carlo (QMC) data  and data from Minimally Entangled Typical Thermal States (METTS) method~\cite{chalopin2026observation}.

\begin{figure}
    \centering
    \includegraphics[width=\columnwidth]{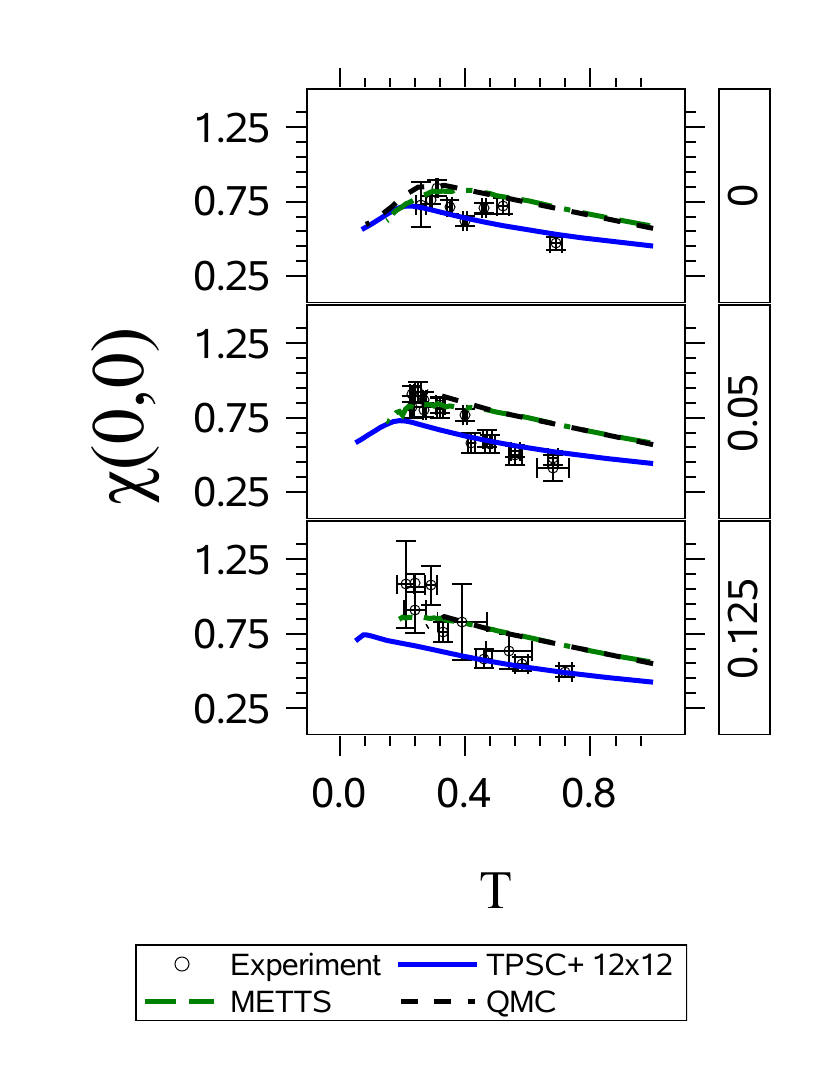} 
    \caption{Temperature dependence of the uniform spin susceptibility $\chi_{sp}(0,0)$ for several dopings ($\delta=0, 0.05, 0.125$) at $U = 6.5$. Solid lines represent results obtained using the TPSC+ approach on a $12 \times 12$ lattice. Symbols correspond to ultracold-atom experimental data, while dashed lines represent results from quantum Monte Carlo (QMC) and METTS methods for systems of comparable size~\cite{chalopin2026observation}. The TPSC+ results are in good qualitative agreement with both experiment and essentially exact numerical methods. All approaches display a maximum in the temperature dependence of the uniform spin susceptibility, a hallmark of the pseudogap. In TPSC+, this maximum occurs at a lower temperature than in the numerical methods and experiments, suggesting that TPSC+ somewhat underestimates the effect of AFM fluctuations at this interaction strength.}
    \label{fig:Chisp00_vs_T_cold_atoms_U_all_Nkx12}
\end{figure}

\fref{fig:Chisp00_vs_T_cold_atoms_U_all_Nkx12} shows the temperature dependence of the uniform spin susceptibility $\chi_{sp}(0,0)$ for several dopings ($\delta=0, 0.05, 0.125$) at $U = 6.5$. The TPSC+ results are in good qualitative agreement with both experimental data and essentially exact numerical methods for finite systems: QMC and METTS. All approaches exhibit a maximum in the temperature dependence of the uniform spin susceptibility-a characteristic hallmark of the pseudogap. In TPSC+, this maximum occurs at a lower temperature than in the numerical methods and experiments, indicating that TPSC+ somewhat underestimates the influence of AFM fluctuations at this interaction strength.

\begin{figure}
    \centering
    \includegraphics[width=\columnwidth]{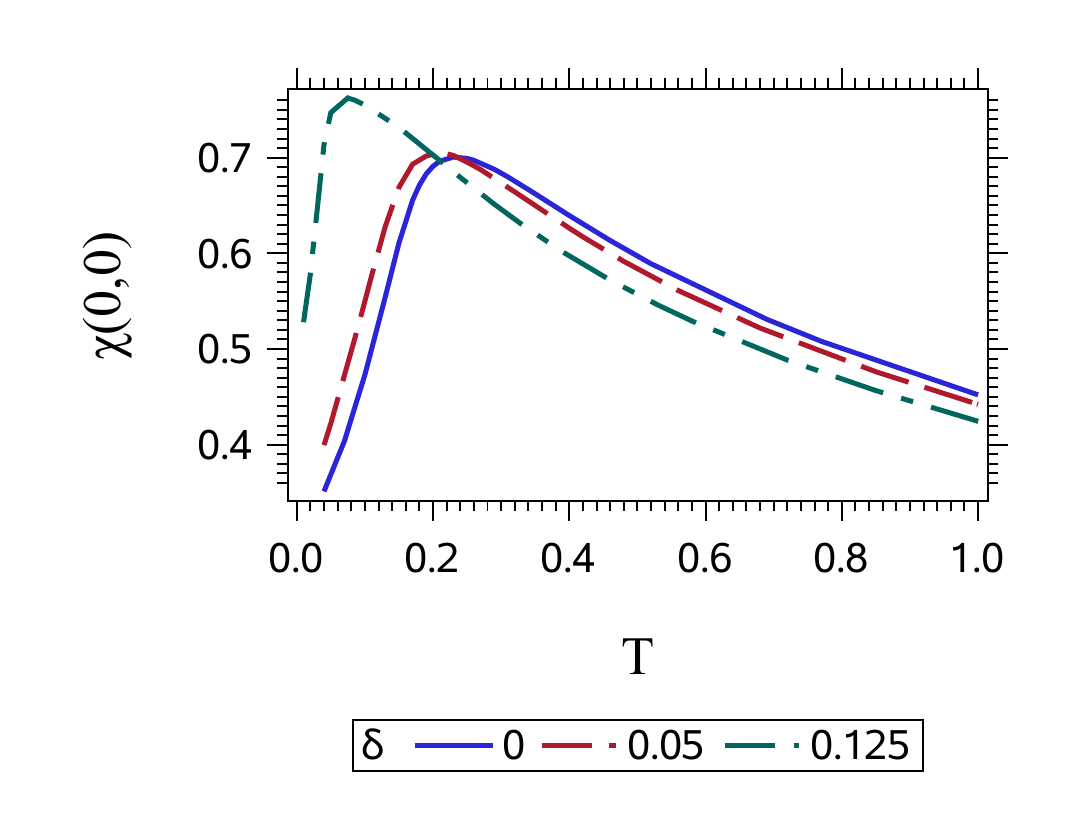} 
    \caption{TPSC+ temperature dependence of the uniform spin susceptibility $\chi_{sp}(0,0)$ for the same dopings and interaction strength as in \fref{fig:Chisp00_vs_T_cold_atoms_U_all_Nkx12}, but calculated in the thermodynamic limit ($\delta=0, 0.05, 0.125$, $U = 6.5$). The maximum in the temperature dependence of the uniform susceptibility is well-defined and shifts to lower temperatures as doping increases.}
    \label{fig:Chisp00_vs_T_cold_atoms_U_6_5_TD}
\end{figure}

The TPSC+ calculations presented above were performed on a $12 \times 12$ lattice, which is comparable to the lattice sizes in ultracold atom experiments. We also investigated finite-size effects in $\chi_{sp}(0,0)(T)$. \fref{fig:Chisp00_vs_T_cold_atoms_U_6_5_TD} shows the TPSC+ results for $\chi_{sp}(0,0)(T)$ obtained in the thermodynamic limit for the same interaction strength and dopings as those used previously. A comparison with \fref{fig:Chisp00_vs_T_cold_atoms_U_all_Nkx12} reveals that while finite-size effects are noticeable at low temperatures, they do not alter the primary findings. The maximum in the temperature dependence of the uniform susceptibility remains well-pronounced and shifts toward lower temperatures as doping increases.

 \section{The evolution of single-particle properties with doping and relation to the maximum in $\kappa(\delta)$ and $\chi_{sp}(0,0)(\delta)$}
\label{sec:N_om_and_A_k_om_evolution}

We consider the cases of commensurate and incommensurate spin fluctuations. The last subsection illustrates the evolution of the spectral function as various momenta cross the Fermi level with changes in doping. 

\subsection{The case of nearly commensurate spin fluctuations} 

 Let us now consider how the maximums in $\kappa(\delta)$ and $\chi_{sp}(0,0)(\delta)$ are related to the evolution of the density of states $N(\omega)$ with doping and to the crossover from the pseudogapped electronic liquid to the Fermi liquid. Using the density of states, the implicit equation for the dependence of the filling $n = 1 - \delta$ on the chemical potential $\mu$ has the well-known form:

\begin{equation}
    \int f(\omega) N(\omega,\mu) d \omega=n.
		\label{eq:mu_equation}
\end{equation}
 At low temperatures, the Fermi function is close to a step function, and the integral in \eref{eq:mu_equation} is primarily over the occupied states, $\omega < 0$.

To put things into perspective, let us start with the evolution of the density of states with doping in the non-interacting case. In this case, and at half-filling, the Fermi surface (FS) is a square, and the density of states exhibits a logarithmic peak at $\omega = 0$ due to the presence of the Van Hove antinodal point $\mathbf{k} = (\pi, 0)$ on the FS. As the system is doped, the FS transforms into an oval within the square; the point $\mathbf{k} = (\pi, 0)$ lies outside of the FS, and the peak in $N(\omega)$ shifts to positive, unoccupied states, $\omega_{peak} > 0$.

This picture does not change qualitatively in the presence of a finite interaction $U > 0$, as long as the system remains a Fermi liquid. In this regime, Luttinger's theorem fixes the area of the FS, its shape changes little, and the Van Hove point remains outside the FS. Consequently, the peak in the density of states stays at positive energies, $\omega_{peak} > 0$.

\begin{figure}
    \centering
    \includegraphics[width=\columnwidth]{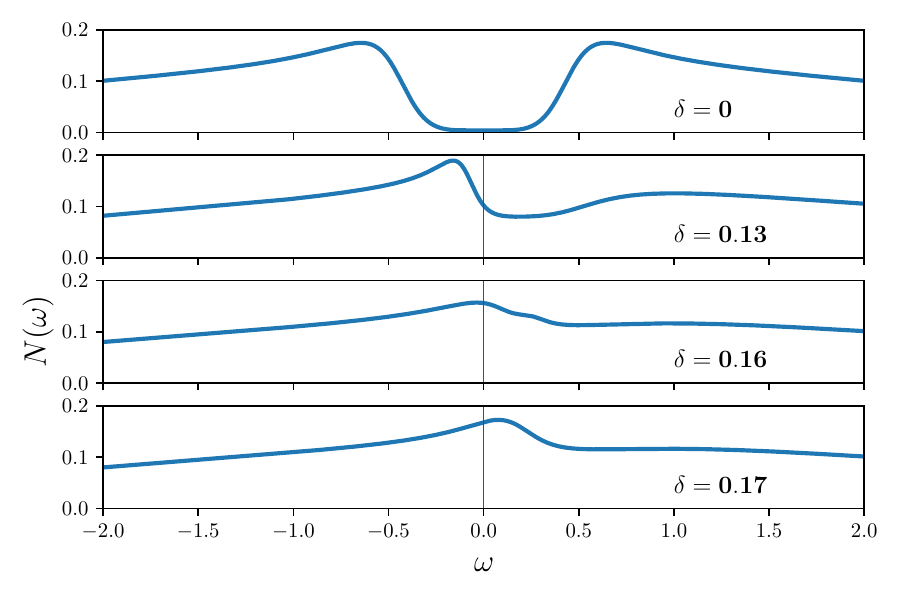} 
    \caption{Evolution of the density of states with doping for $U = 7$ and $T = 0.0714$. For these parameters, a well-defined pseudogap is present at half-filling ($\delta = 0$). As doping increases, the peak corresponding to the precursor of the lower AFM band moves closer to $\omega = 0$ and crosses it at $\delta = 0.16$. This value is very close to $\delta_{max} = 0.15$, at which the compressibility $\kappa(\delta)$ reaches its maximum.}  
    \label{fig:DSw_vs_delta_U_7_T_0_0714}
\end{figure}

However, this picture changes qualitatively in the presence of the pseudogap and the precursors of the AFM bands, as shown in \fref{fig:DSw_vs_delta_U_7_T_0_0714}. In this figure, we show the evolution of the density of states with doping for $U = 7$ and $T = 0.0714$. For these parameters, a well-developed pseudogap is present in the density of states at half-filling. Upon doping, the precursor of the lower AFM band moves toward $\omega = 0$. Because this is where $N(\omega)$ changes shape rapidly, the doping at which this peak crosses the line $\omega = 0$ nearly coincides with the doping $\delta_{max}$ at which the compressibility $\kappa(\delta)$ and uniform magnetic susceptibility $\chi_{sp}(0,0)(\delta)$ reach their maximum. Specifically, $\delta_{max} = 0.15$, and the peak in $N(\omega)$ crosses the line $\omega = 0$ at ${\delta_{peak}|}_{\omega=0} = 0.16$. We found the same relationship at the weak-to-intermediate interaction $U = 3.69$, with the entire picture shifted to smaller doping: $\delta_{max} = 0.1$ and $\delta_{peak}|_{\omega=0} = 0.11$.

\begin{figure}
    \centering
    \includegraphics[width=\columnwidth]{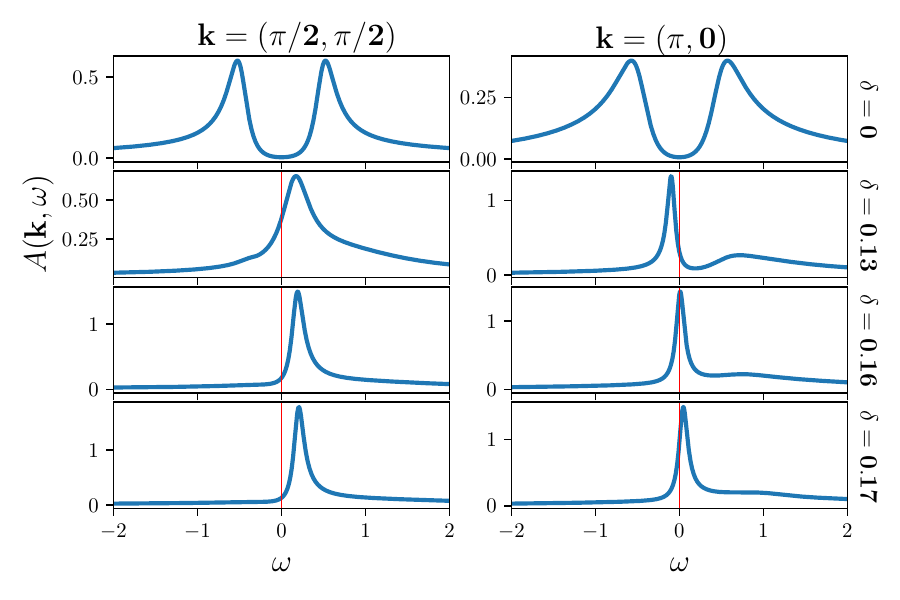} 
    \caption{Evolution of the spectral function $A(\mathbf{k},\omega)$ with doping for $U = 7$ and $T = 0.0714$. For these parameters, a well-defined pseudogap is present at half-filling ($\delta = 0$) for both the nodal and antinodal momentum points. As doping increases, the peak corresponding to the precursor of the lower AFM band moves closer to $\omega = 0$ for both nodal and antinodal points. At doping $\delta = 0.13$, the peak at the nodal point has already shifted to positive energies, while the peak at the antinodal point remains at negative energies. This indicates the persistence of a pseudogap at the antinodal point and the emergence of a Fermi arc near the nodal point, similar to observations in high-$T_c$ materials. The precursor of the lower AFM band at the antinodal point crosses $\omega = 0$ at $\delta = 0.16$, which is the same doping as in the density-of-states case and very close to $\delta_{max} = 0.15$.}
    \label{fig:Akw_by_doping_U_7_T_0_0714}
\end{figure}

 Let us now examine the evolution of the spectral function $A(\mathbf{k},\omega)$ with doping. \fref{fig:Akw_by_doping_U_7_T_0_0714} shows the evolution of $A(\mathbf{k},\omega)$ for the nodal point $\mathbf{k}_N = (\pi/2, \pi/2)$ and the antinodal (Van Hove) point $\mathbf{k}_{AN} = (\pi, 0)$. The parameters are the same as those used in the figure for the density of states above. Under these conditions, a well-developed pseudogap exists for both momentum points at half filling.

As the system is doped, the lower-band precursor of the AFM bands moves toward zero energy. For $\delta = 0.13$, the peak at the nodal point has already shifted to positive energies, while the peak at the antinodal point remains at negative energies. This indicates the persistence of a pseudogap at the antinodal point and the emergence of a Fermi arc near the nodal point, similar to observations in high-$T_c$ materials.

The precursor of the lower AFM band at the antinodal point crosses the line $\omega = 0$ at the same doping, $\delta = 0.16$, where the peak in the density of states also crosses zero. As noted above, this doping is practically the same as $\delta_{max} = 0.15$. We have also performed a similar analysis of $A(\mathbf{k},\omega)$ in the weak-to-intermediate interaction regime, $U = 3.69$, and found qualitatively similar results. The differences are primarily quantitative, with a smaller $\delta_{max} = 0.15$ and a somewhat larger correlation length in the weak-interaction case.

As we pointed out earlier, the maximum of $\chi_{sp}(0,0)(\delta)$ is at higher temperatures than the maximum of $\kappa(\delta) = \partial n / \partial \mu$. For example, for $U = 7$, there is a maximum in $\chi_{sp}(0,0)(\delta)$ at $T = 0.15$ (\fref{fig:chisp00_vs_delta_by_U_T}), but no corresponding maximum in $\kappa(\delta) = \partial n / \partial \mu$ (\fref{fig:Compes_k_by_U_T_inf_Lattice}).

\begin{figure}
    \centering
    \includegraphics[width=\columnwidth]{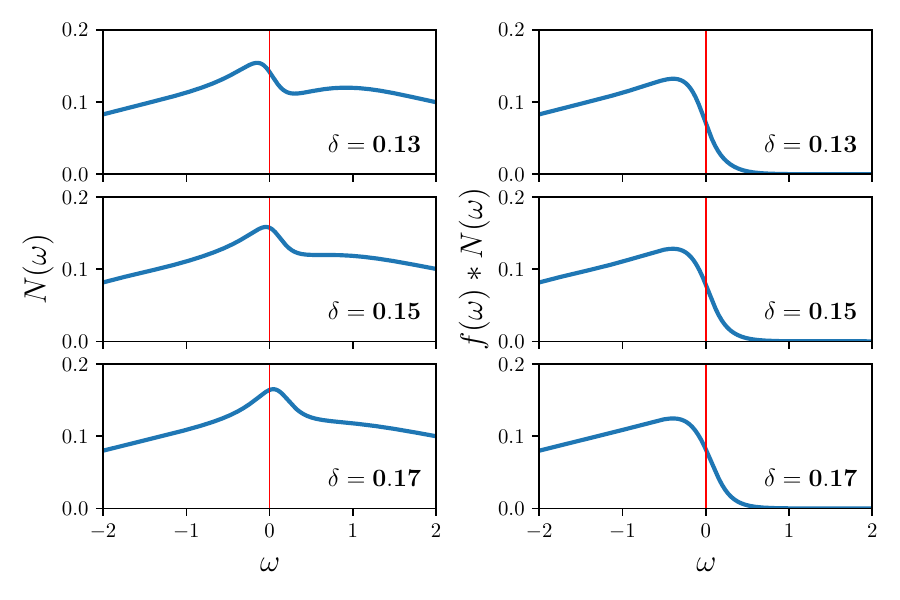} 
   \caption{Evolution of the density of states $N(\omega)$ (left panel) and the product $f(\omega) N(\omega)$ (right panel) with doping. The parameters are $U=7$ and $T=0.15$. While the peak in $N(\omega)$ crosses the line $\omega = 0$ at $\delta = 0.15$, the maximum of the curve $f(\omega) N(\omega)$ does not cross $\omega = 0$. This occurs because the width of the Fermi function $f(\omega)$ at $T=0.15$ is comparable to the width of the peak in the density of states.}
    \label{fig:fDSw_vs_DSw_vs_delta_U_7_T_0_15}
\end{figure}

To understand this difference, note that the implicit equation for the chemical potential, \eref{eq:mu_equation}, depends not only on the density of states $N(\omega)$ but also on the product $f(\omega) N(\omega)$. \fref{fig:fDSw_vs_DSw_vs_delta_U_7_T_0_15} shows the evolution of both the density of states $N(\omega)$ (left panel) and the product $f(\omega)N(\omega)$ (right panel). We see that $N(\omega)$ crosses the line $\omega = 0$ at $\delta_{max} \approx 0.15$, which coincides with the maximum in $\chi_{sp}(0,0)(\delta)$. By contrast, the product $f(\omega)N(\omega)$ changes very little with doping. This occurs because the width of the Fermi function at $T = 0.15$ is comparable to the width of the peak in $N(\omega)$, and, as a result, the Fermi function smears out the peak in $N(\omega)$. Consequently, there is no maximum in $\kappa(\delta) = \partial n / \partial \mu$ for $T = 0.15$.   

\subsection{The case of strongly incommensurate spin fluctuations}
\label{sec:incommensurate_case} 

 As the temperature decreases at a given doping, the antiferromagnetic fluctuations become incommensurate and, instead of a single peak, the spin susceptibility develops four symmetry-related peaks of equal strength~\cite{Schulz_incommensurate:1990}. These peaks are located at the momentum values $Q_1 = (\pi,(1-\alpha)\pi)$, $Q_2 = (\pi,(1+\alpha)\pi)$, $Q_3 = ((1-\alpha)\pi,\pi)$, and $Q_4 = ((1+\alpha)\pi,\pi)$, where $\alpha < 1$.

	Let us recall that in the commensurate case, when $Q = (\pi,\pi)$, the self-energy $\Sigma(\mathbf{k}, \omega)$ develops a quasi-singularity at the frequency $\omega = \tilde{\epsilon}_{\mathbf{k}+\mathbf{Q}}^{(1)}$, $\tilde{\epsilon}_\mathbf{k}^{(1)}=\epsilon_{\mathbf{k}}-\mu^{(1)}$ \cite{Vilk1996,vilk1997shadow}. Since we now have four peaks in the incommensurate case, this leads to up to four quasi-singularities in $\Sigma_(\mathbf{k},\omega)$ located at
\begin{equation}
\omega_i = \tilde{\epsilon}^{(1)}_{\mathbf{k} + \mathbf{Q}_i}.
\end{equation}

 For the nodal and antinodal points, there are only two distinct values of $\omega_i$ because of symmetry. For antinodal point $\mathbf{k}_{AN}=(\pi,0))$ we have  
 $\omega_1=\omega_3$ and $\omega_2=\omega_4$ , while for nodal point $\mathbf{k}_{N}=(\pi/2,\pi/2))$ we have $\omega_1=\omega_2$ and $\omega_3=\omega_4$. The point $\mathbf{k} = (3\pi/4,\pi/4)$ exhibits an even richer structure, with more distinct quasi-singularities.

\begin{figure}
    \centering
    \includegraphics[width=\columnwidth]{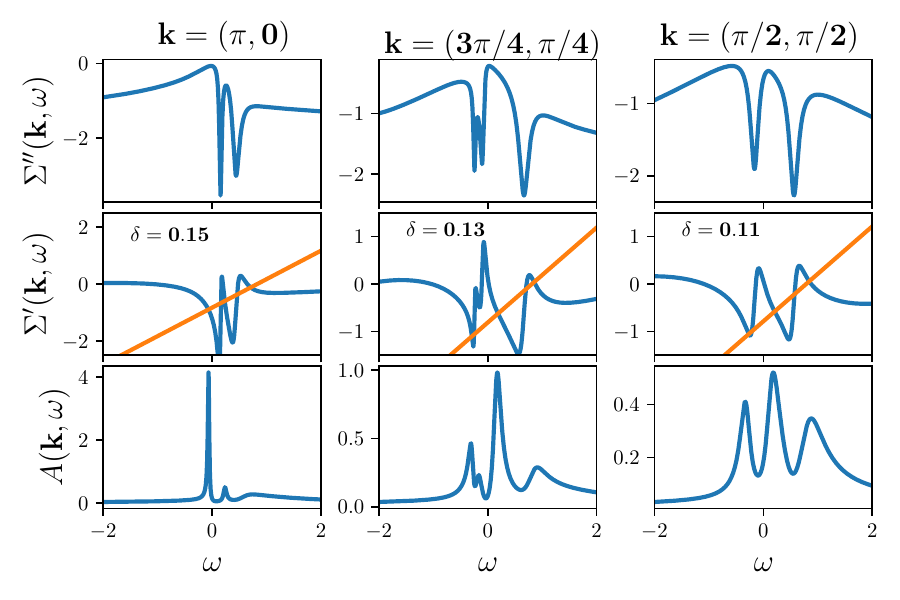} 
   \caption{Effect of the incommensurate spin fluctuations on the single-particle spectra for three momentum values: $\mathbf{k} = (\pi,0)$, $\mathbf{k} = (3\pi/8,\pi/8)$, and $\mathbf{k} = (\pi/2,\pi/2)$.The four peaks in the spin susceptibility lead to multiple quasi-singularities in the self-energy: negative peaks in $\Sigma''(\mathbf{k},\omega)$ and quasi-discontinuities in $\Sigma'(\mathbf{k},\omega)$. As a result, the yellow line $\omega - \epsilon(\mathbf{k}) + \mu$ intercepts $\Sigma'(\mathbf{k},\omega)$ at multiple points, which correspond to the poles of the Green function. This, in turn, leads to more than two peaks in the spectral function $A(\mathbf{k},\omega)$.} 
    \label{fig:Im_S_Re_S_Akw_U_7_incomens_three_points_AN_N1_N}
\end{figure}

\fref{fig:Im_S_Re_S_Akw_U_7_incomens_three_points_AN_N1_N} clearly shows the presence of these singularities and their impact on the spectral function $A_(\mathbf{k},\omega)$. Specifically, the imaginary part of the self-energy, $\Sigma''(\mathbf{k},\omega)$, displays multiple negative peaks at the frequencies $\omega_i$. In the real part, $\Sigma'(\mathbf{k},\omega)$, we observe quasi-discontinuities at the same $\omega_i$. The presence of these discontinuities leads to more than two solutions of the equation for the poles of the Green function:

\begin{equation}
 \omega-\epsilon(\mathbf{k})-\Sigma'(\mathbf{k},\omega) +\mu=0
\label{eq:pols_of_G}
\end{equation} 

 This is illustrated graphically in the middle panel of \fref{fig:Im_S_Re_S_Akw_U_7_incomens_three_points_AN_N1_N}, where we show the intersection of the line $\omega - \epsilon(\mathbf{k}) + \mu$ with $\Sigma'(\mathbf{k},\omega)$. Consequently, in the spectral function $A(\mathbf{k},\omega)$ we see three peaks for both the nodal and antinodal points, and four peaks for the point $\mathbf{k} = (3\pi/4,\pi/4)$. This behavior is qualitatively different from the pseudogap in the commensurate case, in which only two precursors of the AFM band exist. Note that for the point $\mathbf{k} = (3\pi/4,\pi/4)$, two of the four precursors of the SDW bands lie in the occupied-state region $\omega < 0$ and thus can be observed in an ARPES-type experimental probe. Also, note that the incommensurate effect on the spectral function is stronger for the nodal point than for the antinodal point. This is because the singularities are more widely separated in the nodal case:
$|\omega_1(\mathbf{k}_{N}) - \omega_3(\mathbf{k}_{N})| \sim (\alpha \pi) \gg |\omega_1(\mathbf{k}_{AN}) - \omega_2(\mathbf{k}_{AN})| \sim (\alpha \pi)^2 $.

\begin{figure}
    \centering
    \includegraphics[width=\columnwidth]{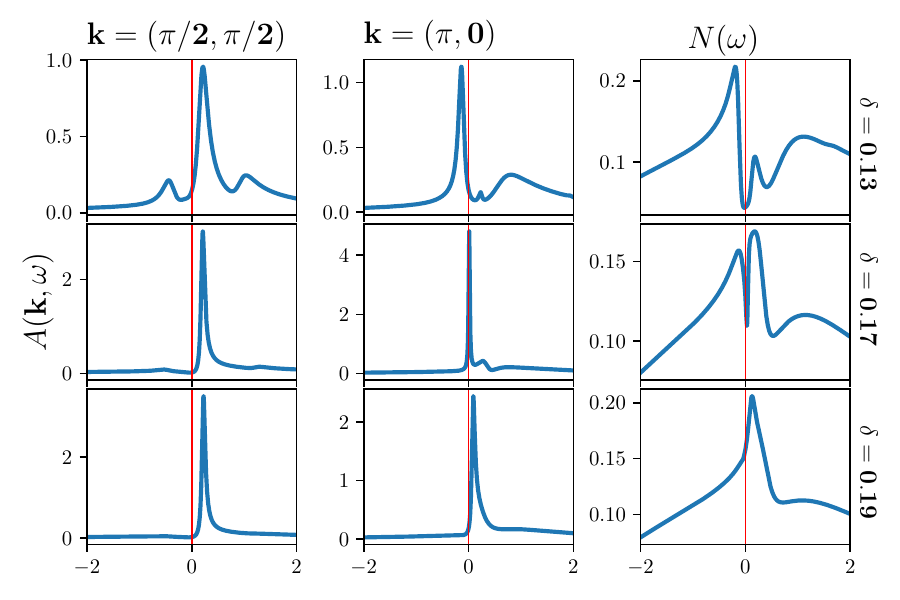} 
   \caption{Evolution of the single-particle spectra with doping when the critical spin fluctuations are incommensurate. The left and middle panels show the spectral function $A(\mathbf{k},\omega)$ for the nodal and antinodal points, respectively. The right panel shows the density of states $N(\omega)$. The temperature is $T=0.025$. As doping increases, the weight of the nodal peak at $\omega < 0$ is transferred to the middle peak at $\omega > 0$. This corresponds to the initial rise in $\kappa(\delta)$ and $\chi_{sp}(0,0)(\delta)$ with doping (before the small plateau). The subsequent increase in these quantities is driven by the evolution of the lower peak at the antinodal point. When this peak crosses the line $\omega = 0$, both $\kappa(\delta)$ and $\chi_{sp}(0,0)(\delta)$ reach their maximum and then decrease rapidly as the antinodal peak moves away from $\omega = 0$. In the density of states, these changes in the spectral function correspond to a transfer of weight from the lower peak at $\omega < 0$ to the middle peak at $\omega > 0$} 
    \label{fig:Akw_N_AN_DS_U_7_T_0_025_incomens}
\end{figure}

  Let us now examine how the evolution of the single-particle spectra with doping leads to the rather complex structure of the maxima in both the compressibility and the uniform magnetic susceptibility. In \fref{fig:Akw_N_AN_DS_U_7_T_0_025_incomens}, the left and middle panels show the spectral function $A(\mathbf{k},\omega)$ for the nodal and antinodal points, respectively, while the right panel shows the density of states $N(\omega)$. As doping increases, the weight of the nodal peak at $\omega < 0$ is transferred to the middle peak at $\omega > 0$. This accounts for the initial rise in $\kappa(\delta)$ and $\chi_{sp}(0,0)(\delta)$ with doping, just before the small plateau. The subsequent increase in these quantities is driven by the evolution of the lower peak at the antinodal point. When this peak crosses the line $\omega = 0$, both $\kappa(\delta)$ and $\chi_{sp}(0,0)(\delta)$ reach their maximum and then decline rapidly as the antinodal peak moves away from $\omega = 0$. In the density of states, these changes in the spectral function correspond to a transfer of weight from the lower peak at $\omega < 0$ to the middle peak at $\omega > 0$.

 In the next \secref{sec:EDC_curves}, we examine the pseudogap at the nodal and antinodal points in more detail by analyzing energy distribution curves (EDCs) for different momenta.
 
\subsection{Evolution of the spectral function $A_{\mathbf{k}}(\omega)$ as momentum crosses the Fermi surface (EDC curves). } 
\label{sec:EDC_curves}

\begin{figure}
    \centering
    \includegraphics[width=\columnwidth]{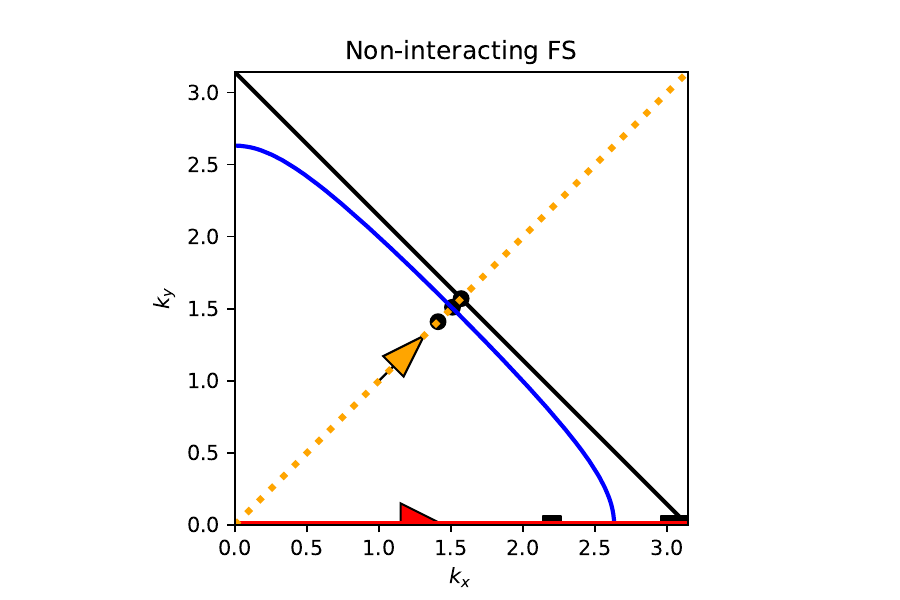} 
   \caption{Part of the Brillouin zone and two of the directions used in \fref{fig:Akw_by_k_U_7_T_0_0714}. The diagonal direction $(0,0)-(\pi/,\pi/2)$ is in orange and the direction  $(0,0)-(\pi,0)$ is in red. Circle and square symbols show momentum points presented in EDC curves. Blue line indicates the non-interacting Fermi surface for doping $\delta=0.13$}
    \label{fig:FS_0_delta_0_13}
\end{figure}

\begin{figure}
    \centering
    \includegraphics[width=\columnwidth]{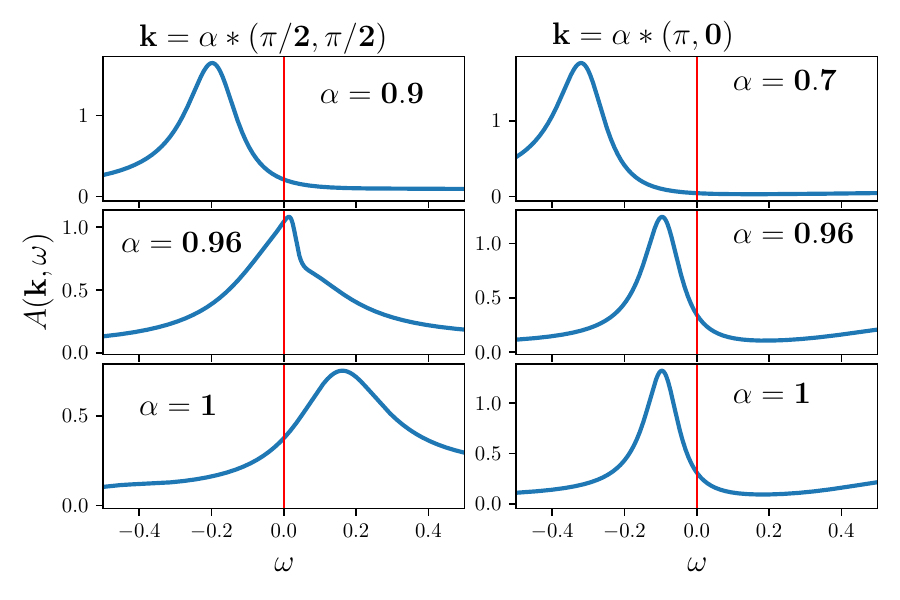} 
   \caption{Evolution of the spectral function $A_{\mathbf{k}}(\omega)$ as momentum crosses the Fermi surface (EDC curves) along two directions defined in \fref{fig:FS_0_delta_0_13}. The parameters are: $U = 7$, $T = 0.0714$, and $\delta=0.13$ ($\xi \approx 4 $). The left panel shows the evolution of $A_{\mathbf{k}}(\omega)$ as $\mathbf{k}$ moves in the diagonal direction $(0,0)-(\pi/2,\pi/2)$, while the right panel shows the evolution of $A_{\mathbf{k}}(\omega)$ in the direction $(0,0)-(\pi,0)$. At this temperature, the EDC curves along the diagonal direction clearly exhibit Fermi-liquid behavior. As momentum approaches and passes through the Fermi surface, the quasiparticle peak crosses zero energy in a typical metallic manner. By contrast, the EDC curves along the $(0,0)$–$(\pi,0)$ direction behave differently: the quasiparticle peak stops at a finite energy and remains nearly stationary, indicating a very flat band ($\omega_{\text{peak}} \propto (\mathbf{k}_{AN})^4$). This is a characteristic signature of the precursor to the lower AFM band near the antinodal point $\mathbf{k}_{AN} = (\pi,0)$. }
    \label{fig:Akw_by_k_U_7_T_0_0714}
\end{figure}

\fref{fig:FS_0_delta_0_13} shows part of the Brillouin zone and two directions used in the \fref{fig:Akw_by_k_U_7_T_0_0714}. Details are in the figure caption.

\fref{fig:Akw_by_k_U_7_T_0_0714} shows the evolution of the spectral function $A_{\mathbf{k}}(\omega)$ as momentum crosses the Fermi surface (EDC curves) along two directions defined in \fref{fig:FS_0_delta_0_13}. The left panel shows the evolution of $A_{\mathbf{k}}(\omega)$ as $\mathbf{k}$ moves in the diagonal direction $(0,0)-(\pi/2,\pi/2)$, while the right panel shows the evolution of $A_{\mathbf{k}}(\omega)$ in the direction $(0,0)-(\pi,0)$. The figure clearly demonstrates the absence of a pseudogap in the first (diagonal) direction and the presence of a pseudogap in the second direction. A more detailed description is given in the figure caption.

  As the temperature is lowered further, the pseudogap spreads from the antinodal (Van Hove) point to regular points on the Fermi surface, leading to a shrinking of the Fermi arc with decreasing temperature. This behavior is driven by the increase of the Fermi velocity $v_F$ as one moves along the Fermi surface from the antinodal to the nodal point. Consequently, the pseudogap criterion $\xi > \xi_{dB\_th} = v_F/(\pi T)$ is the most difficult to satisfy at the nodal point.

In a strictly two-dimensional system, the pseudogap eventually reaches the nodal point at very low temperature, $T = 0.025$, where the correlation length becomes large, $\xi \approx 150$. In a real quasi-two-dimensional system, however, a phase transition to the ordered SDW state may occur before this happens.

\begin{figure}
    \centering
    \includegraphics[width=\columnwidth]{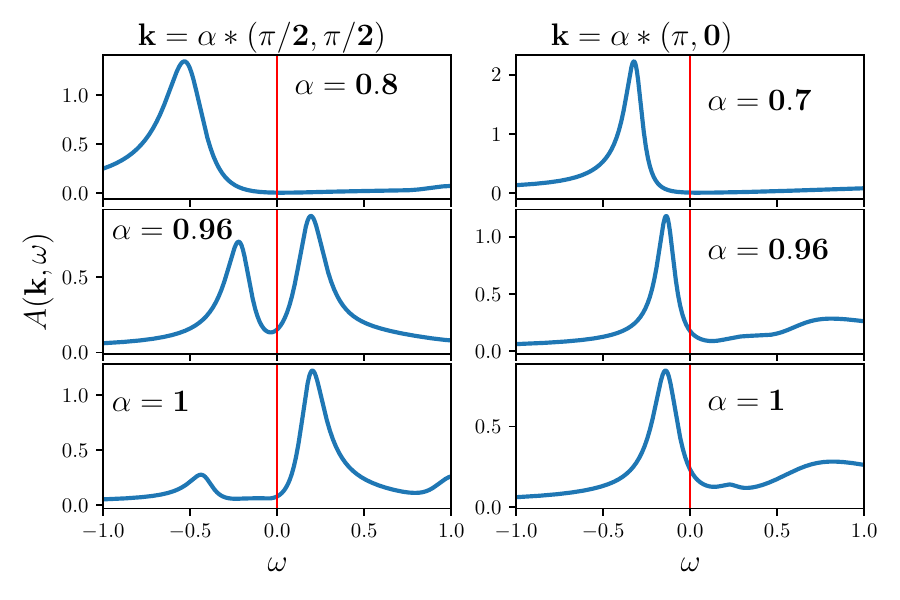} 
  \caption{EDC curves at the same doping and along the same directions as in \fref{fig:Akw_by_k_U_7_T_0_0714}, but at a significantly lower temperature, $T = 0.025$ ($\xi \approx 150$). At this temperature, the EDC curves along the diagonal direction are qualitatively different from those at $T = 0.0714$, clearly exhibiting a pseudogap. As momentum approaches the Fermi surface, the quasiparticle peak first moves toward zero energy, then stops at a finite energy and reverses direction. This is a characteristic signature of the precursor to the lower SDW band. In contrast, the EDC curves along the $(0,0)$–$(\pi,0)$ direction show behavior similar to that at the higher temperature $T = 0.0714$: the peak positions remain nearly unchanged, but the peaks become significantly sharper.}
    \label{fig:Akw_by_k_U_7_T_0_0250}
\end{figure}

\fref{fig:Akw_by_k_U_7_T_0_0250} shows EDC curves along the same directions as before, but at temperature $T = 0.025$. At this temperature, the EDC curves along the diagonal direction clearly exhibit a pseudogap. As momentum approaches the Fermi surface, the quasiparticle peak first moves toward zero energy but then stops at a finite energy and bounces back. This is a characteristic behavior of the precursor to the lower SDW band. In contrast, the EDC curves along the $(0,0)$–$(\pi,0)$ direction show behavior similar to that at the higher temperature $T = 0.0714$: the peak positions remain nearly unchanged, but the peaks become sharper. Note that the AFM fluctuations become incommensurate at this temperature.

\section{A comparison between $\partial n/\partial \mu$ calculated using the second-level self-energy and $\chi_{ch}(0,0)$ calculated using $U_{ch}$ at the first level}
\label{sec:dn_over_dm_vs_chich00}

\begin{figure}
    \centering
    \includegraphics[width=\columnwidth]{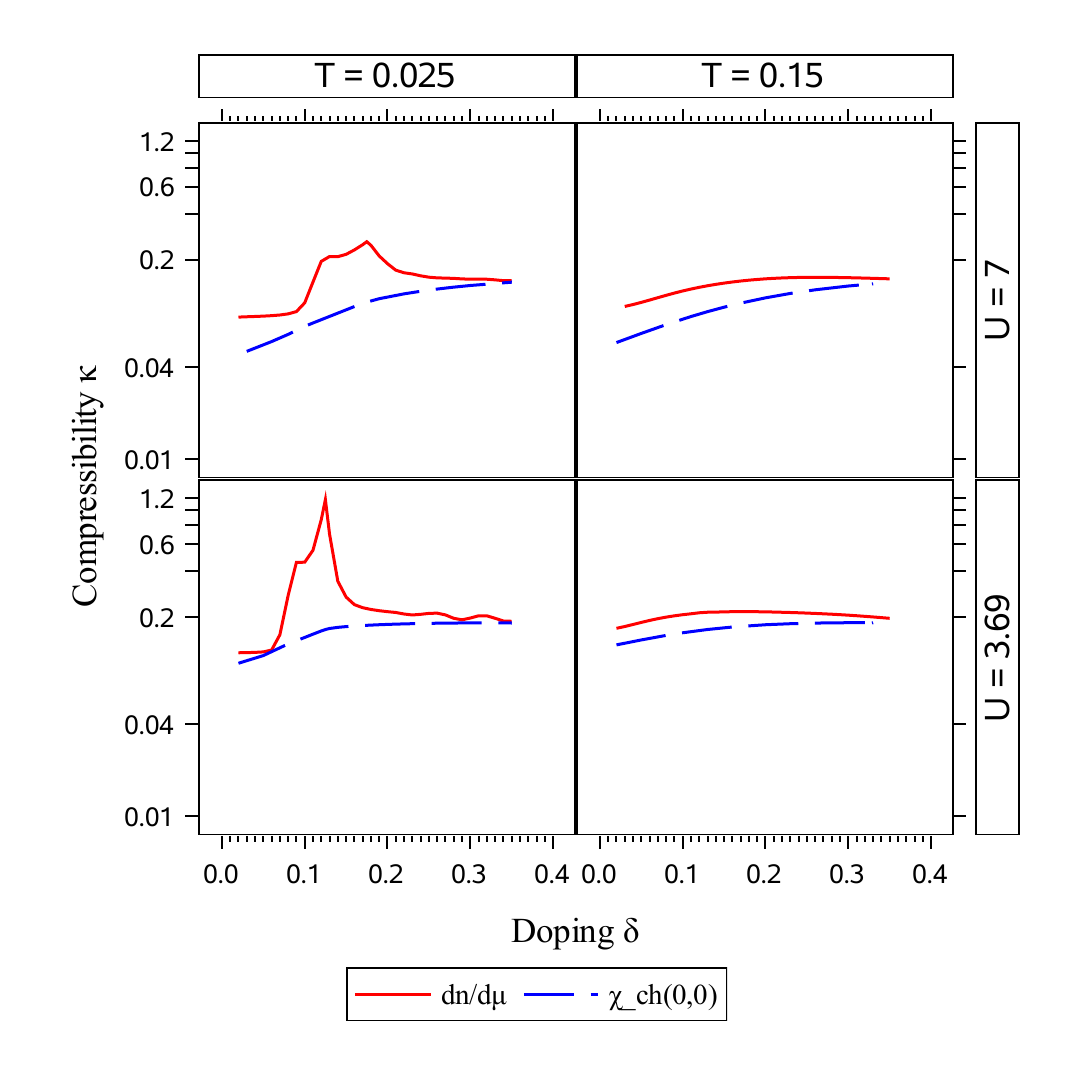} 
  \caption{Comparison between $\partial n/\partial \mu$ calculated using the second-level self-energy and $\chi_{ch}(0,0)$ calculated using a frequency and momentum independent $U_{ch}$ at the first level. In the exact theory, $\partial n/\partial \mu = \chi_{ch}(0,0)$ holds identically. In TPSC+, at the relatively high temperature $T = 0.15$, both quantities take similar values and show no well-defined maximum as a function of doping $\delta$. By contrast, at the much lower temperature $T = 0.025$, the results remain similar at high doping in the Fermi-liquid but differ qualitatively for low to intermediate doping in the pseudogap regime. This indicates that the assumption of a momentum- and frequency-independent charge interaction, $U_{ch} = \text{const}$, becomes inadequate at very low temperatures deep in the pseudogap regime.}
    \label{fig:der_n_over_mu_vs_chich00_T_0_15_0_025}
\end{figure}

In the exact theory, $\partial n/\partial \mu$ must be equal to $\chi_{ch}(0,0)$. In TPSC+, however, we compute $\partial n/\partial \mu$ using the second-level self-energy \eref{eq:selfEnergy2}, while $\chi_{ch}(0,0)$ is calculated using a momentum and frequency independent $U_{ch}$ at the first level, \eref{eq:chich2} and \eref{eq:sumrule_ch}. The comparison between these two calculations in TPSC+ is shown in \fref{fig:der_n_over_mu_vs_chich00_T_0_15_0_025} for two temperatures, $T = (0.025, 0.15)$, and two interaction strengths, $U = (3.69, 7)$. For large doping, all calculations in \fref{fig:der_n_over_mu_vs_chich00_T_0_15_0_025} give similar results. And they also agree with experiment~\cite{kendrick2025pseudogap}. At the relatively high temperature $T = 0.15$, both quantities take comparable values and show no well-defined maximum as a function of doping $\delta$. By contrast, at the much lower temperature $T = 0.025$, the results remain similar at high doping in the Fermi-liquid, but differ qualitatively at low to intermediate doping in the pseudogap. The mathematical reason for the monotonically increasing $\chi_{ch}(0,0)(\delta)$ is the large and rapidly decreasing $U_{ch}(\delta)$.
 
  As we discussed in \secref{sec:TPSC+}, at the first level of approximation, in TPSC and TPSC+, we assume that the charge vertex is a constant and determine it from the charge sum rule and the Pauli principle, \eref{eq:sumrule_ch} and \eref{eq:Pauli_principle}. This approach works well at relatively high temperatures as compared with available Monte Carlo results~\cite{vilk1994theory,gauvin2023improved}. However, in reality, the irreducible charge vertex is a complex function of three independent momenta $\mathbf{q_1},\mathbf{q_2},\mathbf{q_3}$ and three independent Matsubara frequencies $iq_{n1},ik_{n2},ik_{n3}$. The approximation $U_{ch}=\text{const}$ connected to susceptibility that is irreducible in the charge channel assumes that it is a good approximation to replace the full irreducible charge vertex by a real number that is a sort of average. An improved approach would be to parametrize the charge susceptibility using an RPA-like form but with a simplified momentum- and frequency-dependent effective interaction $U'_{ch}(\mathbf{q},iq_n)$.
	
\begin{equation}
    \chi_{ch}(\mathbf{q},iq_n) = \frac{\chi^{(2)}(\mathbf{q},iq_n)}{1+U'_{ch}(\mathbf{q},iq_n)\chi^{(2)}(\mathbf{q},iq_n)/2}.
    \label{eq:chichqiqn}
\end{equation}

\begin{figure}
    \centering
    \includegraphics[width=\columnwidth]{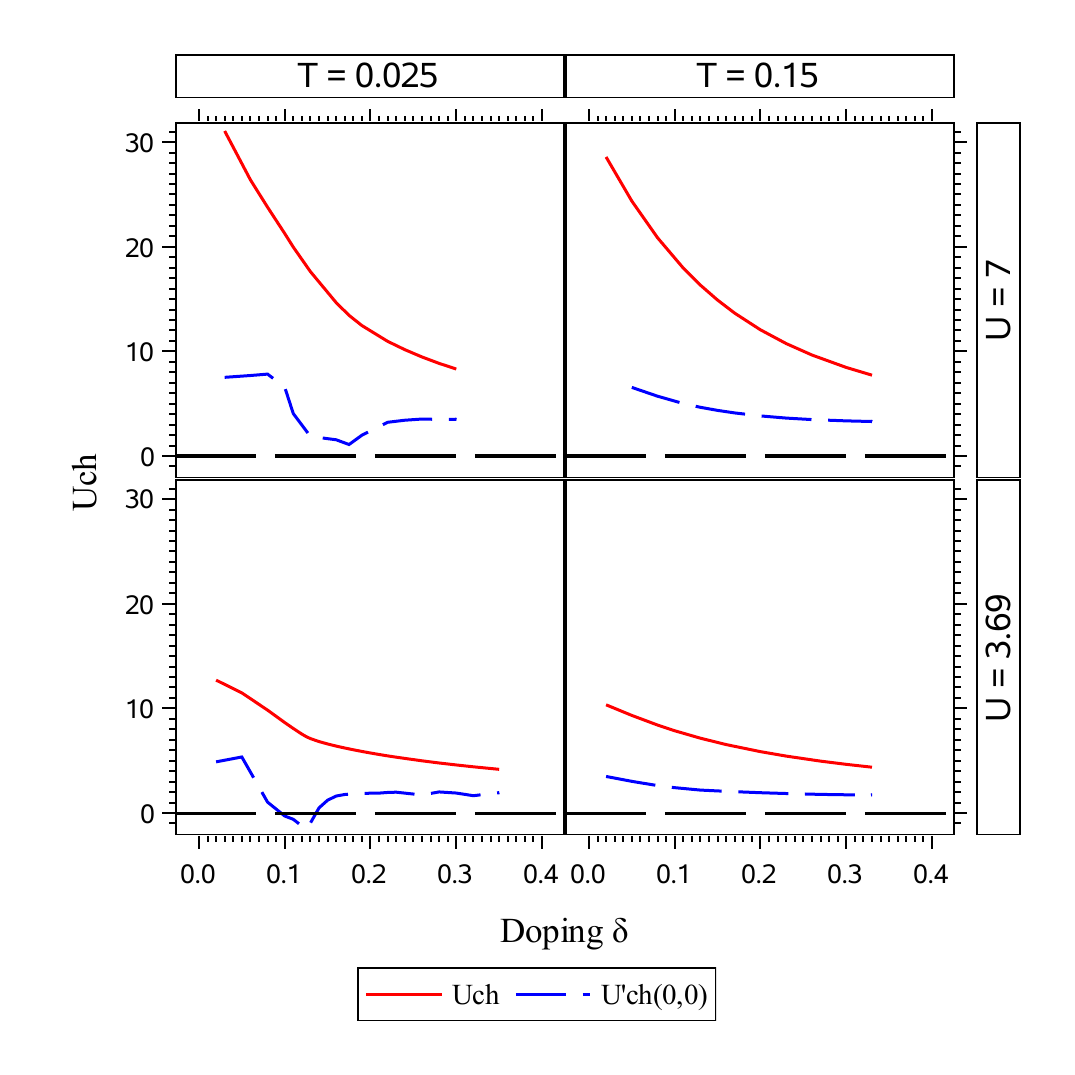} 
  \caption{Comparison between the doping dependence of $U'_{ch}(0,0)$ (see main text) and the average $U_{ch}$ determined at the first level of the TPSC+ approximation using the charge sum rule and the Pauli principle. We see that $U'_{ch}(0,0)$ is significantly smaller than the average $U_{ch}$ from the first-level approximation and, moreover, becomes negative at sufficiently low temperature. }
    \label{fig:Uch_vs_Uch00_T_0_15_0_025}
\end{figure}

 Let us then invert the problem and determine the effective interaction $U'_{ch}(0,0)$ that reproduces $\chi_{ch}(0,0)=\partial n/\partial \mu$. The comparison between $U'_{ch}(0,0)$ and the average $U_{ch}$ is shown in \fref{fig:Uch_vs_Uch00_T_0_15_0_025}. We see that $U'_{ch}(0,0)$ is much smaller than the average $U_{ch}$ obtained in the first level of approximation and, moreover, becomes negative at sufficiently low temperature and some values of $\delta$ and $U=3.69$. When $U'_{ch}(0,0) < 0$, it is very likely that the effective interaction $U'_{ch}(\mathbf{q},0)$ also becomes negative at some finite $q$-value. This opens up the possibility of an SDW-driven incommensurate CDW at very low temperatures deep in the pseudogap regime. Most finite-temperature calculations cannot currently reach such low $T$, which may explain the disconnect between present finite-temperature work and ground-state studies that do find a stripe phase \cite{qin2022hubbard,xu2022stripes,Simkovic_2021}. 

  To quantitatively analyze charge modulations in the pseudogap regime, the TPSC+ approach must be extended to include a momentum-dependent effective interaction $U_{ch}(\mathbf{q})$. We are exploring the feasibility of this extension.
	
\section{Dependence of the correlation length on doping in the weak- and intermediate-interaction limits}
\label{sec:xi_vs_delta_two_U}

\begin{figure}
    \centering
    \includegraphics[width=\columnwidth]{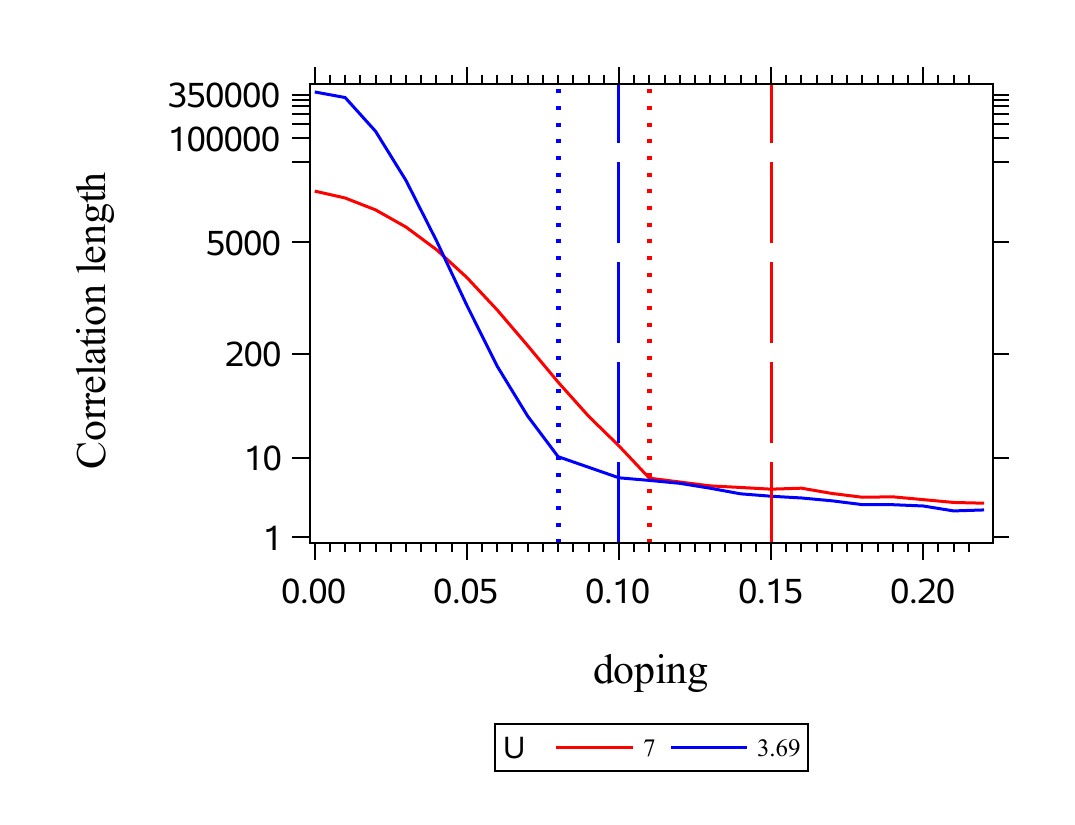} 
    \caption{ Semi-logarithmic plot for the dependence of the correlation length on doping at $T=0.0714$ for two interaction strengths $U=(3.69,7)$. For these $T$ and $U$, the correlation length at half-filling is enormous for both interaction strengths. Any real quasi two-dimensional system would become ordered due to three-dimensional (3D) effects. As doping increases, three regions can be identified: At small doping, the decrease in the correlation length is relatively slight, especially in the case of intermediate interactions. This is the region where one would expect a true antiferromagnetic state in real quasi 2D materials. As doping increases further, the system enters a region with exponentially decreasing $\xi(\delta)$, as indicated by the straight line on the logarithmic scale used in the figure. This region ends at the doping where the Spin Density Wave (SDW) fluctuations become incommensurate. This doping level is indicated by the dotted reference lines of the corresponding color. As doping increases further, the system passes the crossover line at the corresponding $\delta_{max}$ indicated by the dashed line. For both interaction strengths, the crossover occurs when the SDW is incommensurate.}
    \label{fig:xi_vs_delta_T_0_0714_by_U}
\end{figure}

 Let us now examine the dependence of the correlation length on doping in the weak- and intermediate-interaction limits. The correlation length has been computed using the method described in \cite{gauvin2023improved,vilk2024antiferromagnetic}.  The semi-logarithmic plot \fref{fig:xi_vs_delta_T_0_0714_by_U} shows $\xi(\delta)$  for $U = (3.69, 7)$ and $T = 0.0714$. Under these conditions, the correlation length $\xi$ at half-filling is enormous for both interaction strengths. In any real quasi-two-dimensional system, such large values would lead to ordering due to three-dimensional effects.

 As doping increases, three distinct regions can be identified. At small doping, the decrease in the correlation length is relatively slight, particularly in the intermediate-interaction case. This is the regime where one would expect a true AFM state to appear in real quasi-2D materials. As doping increases further, the system enters a region where $\xi$ decreases exponentially, as indicated by the straight-line behavior in \fref{fig:xi_vs_delta_T_0_0714_by_U}. This region ends at the doping level where spin-density-wave (SDW) fluctuations become incommensurate. In \fref{fig:xi_vs_delta_T_0_0714_by_U}, this doping is marked by the dotted reference line of the corresponding color.

 At even higher doping, the system crosses $\delta_{max}$ indicated by the dashed line. Notably, for both interaction strengths, the crossover between pseudogap and correlated Fermi liquid occurs when the SDW fluctuations are already incommensurate.

\section{Discussion}
\label{sec:Discussion}
We discuss in turn the link between our work and the compressibility maximum found along the Widom line in CDFMT~\cite{sordi2012pseudogap}, and the compressibility maximum found with infinite projected entangled pairs~\cite{Sinha_Wietek_2025}.
We end with a discussion of spin-fluctuation-mediated theories of the pseudogap and their possible relevance to cuprate physics.
\subsection{Widom line and compressibility maximum}
\label{sec:Widom}
The CDMFT method used in Ref.~\cite{sordi2012pseudogap} solves exactly a $2 \times 2$ plaquette of interacting electrons, including a hybridization function that represents the retarded self-consistent lattice environment~\cite{Maier_Jarrell_Pruschke_Hettler_2005,Kotliar_Savrasov_Haule_Oudovenko_Parcollet_Marianetti_2006,Tremblay_Kyung_Senechal_2006,Pavarini:852559}.
With this method, it was found in Refs.~\cite{Sordi_Haule_Tremblay_2010,Sordi_Haule_Tremblay_2011} that the first-order Mott transition at half-filling extends to finite doping, where a first-order transition (Sordi transition) separates a pseudogap from a correlated metal.
Above the critical endpoint of that first-order transition at finite doping, various crossover lines are converging together, the Widom line, towards the critical endpoint. 
The Widom-line concept emerged in classical fluids~\cite{Xu_Kumar_Buldyrev_Chen_Poole_Sciortino_Stanley_2005}.
Among the various extrema of thermodynamic quantities along the Widom line, Ref.~\cite{sordi2012pseudogap} found a compressibility maximum at the crossover between the pseudogap and the correlated electron liquid.  

Given that this phenomenon appeared only for values of $U$ large enough that there is a Mott transition at half-filling, the compressibility maximum was seen as a signature of intermediate interactions. 
This is in clear contradiction with our results that show a compressibility maximum for small values of $U$ at low-enough temperature. 
In addition, the compressibility maximum in CDMFT appears at dopings systematically much smaller than those observed in our calculations and in cold-atom experiments. 

Nevertheless, there are similarities between the results. 
Indeed, with the CDMFT method: a)~the small lattice size shows that the compressibility maximum occurs even with a small antiferromagnetic correlation length; b)~the magnitude of the compressibility maximum increases as temperature decreases; c)~the pseudogap is strongest at the antinodal point, needing very small correlation length~\cite{Meixner_Menke_Klett_Heinzelmann_Andergassen_Hansmann_Schafer_2024}; d)~a maximum in the uniform spin susceptibility as a function of doping is also observed~\cite{Sordi_Haule_Tremblay_2011}. 
But, contrary to the present calculation, in CDMFT the spin susceptibility has a maximum at a doping that is necessarily larger than the doping of the maximum in the compressibility. 
The difference between these two dopings can be large, increasing with $U$. 

We argue that for the genuine CDMFT compressibility maximum to be observed, one needs the strong frustration provided by small cluster size or, for example, by a triangular lattice~\cite {Downey_Gingras_Hebert_Charlebois_Tremblay_2024}. 
Indeed, finite-temperature fluctuations of an ordered ground state can, in certain cases, completely hide the Mott-insulator plus superexchange physics taken into account by CDMFT in small clusters~\cite{schafer2015fate, Reymbaut_Boulay_Fratino_Sémon_Wu_Sordi_Tremblay_2020, Meixner_Menke_Klett_Heinzelmann_Andergassen_Hansmann_Schafer_2024, Fratino_AFM:2017}. 
CDMFT is geared for large interactions, large temperatures, and the small correlation lengths provided by frustration. 

Because of the frustration provided by the relatively large next-nearest-neighbor hopping found in several cuprate high-temperature superconductors, many finite-temperature cluster generalizations (CDMFT) (DCA)~\cite{Lichtenstein:2000,Biroli:2002,Jarrell:2001a,Maier_Jarrell_Pruschke_Hettler_2005,Kotliar_Savrasov_Haule_Oudovenko_Parcollet_Marianetti_2006,Tremblay_Kyung_Senechal_2006} of dynamical mean-field theory~\cite{georges1996,georges1992,jarrell1992} give results on the square lattice that compare favorably with the phenomenology of the cuprate pseudogap ~\cite{huscroft2001,macridin2006,kyung2006pseudogap,stanescu2006,Ferrero_Cornaglia_De_Leo_Parcollet_Kotliar_Georges_2009,sakai2009,Vidhyadhiraja:2009,sordi2012pseudogap,SordiSuperconductivityPseudogap:2012,Gull_Millis_2013,SordiResistivity:2013,gull2013,Fratino_Semon_Sordi_Tremblay_2016,Reymbaut:2019,dash2019,Walsh_Sound_2022,Sordi_Specific_heat_2019,Walsh_Sordi_Opalescence_2019,Aguiar_Bragança_Paul_Civelli_2025},

CDMFT calculations for the Hubbard model on a triangular lattice, on the other hand~\cite {Downey_Gingras_Hebert_Charlebois_Tremblay_2024}, offer the possibility of comparisons with experimental results on layered organic materials such as $\kappa$-BEDT layered organic compounds~\cite{oike:2017, yoshimoto:2013,yamamoto:2013}.

It is important to note that a maximum in compressibility was also seen in DCA calculations~\cite{Khatami_Mikelsons_Galanakis_Macridin_Moreno_Scalettar_Jarrell_2010}. 
%
%
%
However, the phase separation found at positive $t'$ is interpreted as a phase separation between a Mott insulator and a metal~\cite{Macridin_Jarrell_Maier_2006,Galanakis_Khatami_Mikelsons_Macridin_Moreno_Browne_Jarrell_2011}, not between a pseudogap metal and a correlated Fermi liquid.
It is claimed that $t'=0$ is a quantum critical point.
\subsection{Compressibility maximum with infinite projected entangled pairs}
Based on tensor network methods, more specifically infinite projected entangled pair states with purification in the thermodynamic limit, Ref.~\cite{Sinha_Wietek_2025} has shown that $\kappa$ develops a broad maximum above the stripe regime near $\delta=0.1$. 
The magnitude of this maximum increases upon cooling, like we found.
The doping where the phenomenon is observed is too large compared with $2\times2$ CDMFT results~\cite{sordi2012pseudogap}. 
However, comparing $\delta_{max}=0.1$ for the compressibility found in that work at $T=0.125$ and $U=10$ with the $T=0.15$ and $T=0.1$ phase diagrams in Fig.~(1) of Ref.~\cite{Simkovic_Rossi_Georges_Ferrero_2024}, we see that this $\delta_{max}=0.1$ is consistent with a simple-minded extrapolation to $U=10$ of the antinodal pseudogap location, in light red, appearing in that phase diagram. 

It is important to note that, because the filling is given by \eref{eq:mu_equation}, any fluctuating order or long-range order that modifies the single-particle spectral weight, as in ~\figref{fig:DSw_vs_delta_U_7_T_0_0714} to \figref{fig:Akw_by_k_U_7_T_0_0250}, will lead to a maximum in the compressibility upon entering or leaving the fluctuating or long-range order.
Even though this signal appears in the $\mathbf{q=0}$ charge sector, the fluctuating or long-range order may reside in any channel, including purely spin channels.
Hence, the compressibility maximum is not tied to weak or strong interactions; while it indicates a softening of charge fluctuations, the leading mechanism may originate from a different channel.

\subsection{Spin-fluctuation-mediated theories of the pseudogap}

Several theories of the pseudogap have been proposed. 
This is not a review paper, so we do not discuss all theories of the pseudogap. 
We focus instead on spin-fluctuation theories of the pseudogap, which is the general context for our approach. 
We even set aside spin-fluctuation mechanisms for d-wave superconductivity, another fascinating subject.

We stress that apart from results from a recent approximate representation of the two-dimensional Hubbard model~\cite{Zhao_Mai_Tenkila_Phillips_2025}, none of the spin-fluctuation approaches, except ours, have yet addressed the question of the compressibility maximum as a function of doping.
Given the physics of the compressibility maximum that we have presented, we expect this phenomenon to be found in many theories that exhibit a pseudogap. 

Early approaches were based on nearly antiferromagnetic Fermi liquids and spin-fermion models~\cite{schmalian_microscopic_1999}. 
Refs.~\cite{Chubukov_Pines_Schmalian_2002} and~\cite{Moriya_Ueda_2000} contain reviews of this early work that does not address the Hubbard model directly, but that, through a Hubbard-Stratonovich decomposition, posits a susceptibility for propagating~\cite{Chubukov_1998} or overdamped~\cite{Millis_Monien_Pines_1990} spin waves and a coupling constant to fermions.
Several existing phenomenological versions of spin-fluctuation theories are developed in Ref.~\cite{Verret_Simard_Charlebois_Senechal_Tremblay_2017}.
Quantum critical behavior of these theories was reviewed in this Ref.~\cite{Abanov_Chubukov_Schmalian_2003}.
These approaches then contain several adjustable parameters that allow successful comparisons with experiments, especially for the most recent versions of spin-fermion theories~\cite{kokkinis2025pseudogap}, which are based on critical thermal spin fluctuations that are precursors to an SDW ground state. 

%
For the near-neighbor hopping two-dimensional Hubbard model, Refs.~\cite{Vilk1996,Vilk1997} demonstrated that critical thermal spin fluctuations can lead to a pseudogap as a precursor to a zero-temperature spin-density wave (SDW) ground state, a phenomenon whose physics is analogous to early work in the context of quasi-one-dimensional systems~\cite{lee1973fluctuation} but based on charge-density wave (CDW) fluctuations instead of SDW.
To demonstrate this phenomenon, it was necessary to develop an approach that satisfies the Hohenberg-Mermin-Wagner theorem~\cite{Mermin:1966,Hohenberg:1967}, the two-particle self-consistent approach (TPSC)~\cite{Vilk1996,Vilk1997}, whose generalization we have applied here. 

TPSC is a non-perturbative method with no direct diagrammatic interpretation.
By contrast, the first spin-fluctuation approaches to the Hubbard model based on diagrammatic expansions are the paramagnon theories~\cite{Enz_1992,Stamp_1985}. 
This kind of approach is based on an RPA calculation of the spin and charge susceptibilities, which are then fed back into the self-energy. 
This clearly cannot explain the pseudogap in the two-dimensional Hubbard model since RPA leads to a phase transition in two dimensions instead of an RC regime. 

The Fluctuation Exchange approximation (FLEX)~\cite{Bickers:1989,Bickers_dwave:1989,Bickers:1991,Bickers_2004} is a more sophisticated approach belonging to the category of conserving approximations. 
Any set of two-particle irreducible skeleton diagrams used to define a Luttinger-Ward functional generates approximations that satisfy conservation laws, but not crossing symmetry (Pauli principle) in general. 
FLEX corresponds to a particular RPA-like set of skeleton diagrams for spin and charge fluctuations. 
Because of the feedback of the self-energy on the spin-susceptibility, this approach satisfies the Mermin-Wagner theorem. 
It leads to a pseudogap in the density of states, but not in the ARPES spectrum~\cite{Moukouri2000}, for reasons that are understood~\cite{Vilk1997}.
Other resummations of infinite sets of diagrams~\cite{kuchinskii_models_1999,Sedrakyan:2010,Ye_Wang_Fernandes_Chubukov_2023,Ye_Chubukov_2023} link the pseudogap to antiferromagnetic fluctuations. 

Here is a small sample of the work done using the functional renormalization group~\cite{ZanchiSchulz:2000,Honerkamp_Salmhofer_Furukawa_Rice_2001,Rohe_Metzner_2005, Hille_Rohe_Honerkamp_Andergassen_2020} appropriate for weak interactions.
This work emphasizes the importance of umklapp processes, in addition to spin fluctuations,  in creating the pseudogap. 
Unfortunately, the renormalized interactions diverge upon entering the renormalized classical regime, forbidding the study of the well-developed pseudogap.

So-called Parquet equations are a diagrammatic reformulation of the Many-Body problem that satisfies crossing symmetry~\cite{de_dominicis_stationary_1964,Bickers_2004}.
In practice, the equations are impossible to solve exactly. The pseudo-potential Parquet approach~\cite{Bickers_Scalapino_1992} was an early attempt to solve these equations, at least approximatively.
It was not possible to reach regimes that went beyond second-order perturbation theory~\cite{Vilk1997}, so again the pseudogap remained out of reach.
Recent progress in numerical solutions of the Parquet equations~\cite{Parquet_Gunnarsson:2016} shows that spin fluctuations dominate and that instabilities appear in the charge channel. 
A new finite-difference parquet method developed for strong interactions~\cite{lihm2025finitedifferenceparquetmethodenhanced} concludes that even in this case, it is enhanced electron-paramagnon scattering that opens a pseudogap.

Monte Carlo summations of Feynman diagrams as a direct method to solve the Hubbard model appear, for example, in these Refs.~\cite{Houcke_Kozik_Prokofev_Svistunov_2010,rossi2018directsamplingselfenergyconnected,Simkovic_2021,Simkovic_Rossi_Georges_Ferrero_2024}. 
Ref.~\cite{Simkovic_Rossi_Georges_Ferrero_2024} qualitatively connected the origin of the pseudogap in both weak and strong interaction limits to spin fluctuations. 
Many recent diagrammatic approaches are based on DMFT vertices to include the effect of Mott physics.
For example, the TRILEX approach~\cite{Ayral_Parcollet_2015} interpolates between strong and weak interactions by decoupling the electron-electron interaction with a Hubbard-Stratonovich transformation and then making a local and self-consistent approximation for the irreducible three-leg vertex using a quantum impurity model in the spirit of DMFT.
Weak signatures of the pseudogap were found in this approach.

More involved four-leg vertices obtained from DMFT have also been used to develop a variety of approaches. 
In particular, to understand the strong-interaction limit, the method of Ref~\cite{Krien_Valli_Chalupa_Capone_Lichtenstein_Toschi_2020} based on the so-called boson exchange parquet~\cite{Bickers_2004} solver for dual fermions~\cite{Rubtsov_Katsnelson_Lichtenstein_2008,Astretsov_Rohringer_Rubtsov_2020} was used in Ref.~\cite{krien2022} to study the pseudogap problem. 
This method, which justifies some aspects of the early spin-fermion theories~\cite{Krien_Lichtenstein_Rohringer_2020}, suggests that when interactions are strong, the effective spin-fermion vertex becomes complex with a large imaginary part.
This leads to a larger dichotomy between the nodal and antinodal parts of the Fermi surface, explaining why the hole-doped cuprates exhibit a pseudogap even when the spin correlation length is small. 
We have argued in this paper that even for modestly strong $U < 7$ interactions, a small spin correlation length suffices to open a pseudogap near the antinodal regions that satisfy $v_F < \gamma/\xi$~\cite{vilk2023criteria}. 
Dual fermions are a strong-interaction approach based on a Grassmann-Hubbard-Stratonovich transformation of the kinetic energy term~\cite{Pairault:2000}. 
Some of the limitations of dual fermions have been discussed~\cite{Gukelberger_Huang_Werner_2015}.

One of the prominent approaches based on vertices obtained from DMFT as a basis for diagrammatic approaches is D$\Gamma$A.
In this approach, spin fluctuations are also at the origin of the pseudogap in the two-dimensional Hubbard model on a square lattice~\cite{Kaufmann_Eckhardt_Pickem_Kitatani_Kauch_Held_2021}.

While one must worry about periodization artefacts~\cite{Verret_Roy_Foley_Charlebois_Sénéchal_Tremblay_2019,Verret_Simard_Charlebois_Senechal_Tremblay_2017}, CDMFT or DCA observe a pseudogap. But identifying the mechanism is usually based on correlating different observables. 
For example, in Refs.~\cite{civelli:2005,Ferrero_Cornaglia_DeLeo_Parcollet_Kotliar_Georges_2009,kyung2006pseudogap,Haule_plaquette_2007,Sordi_Haule_Tremblay_2010,Sordi_Haule_Tremblay_2011,Reymbaut:2019} short-range spin or singlet fluctuations were identified as the origin of the pseudogap. 
In DCA, it is claimed that the pseudogap comes from spin fluctuations~\cite{macridin2006} and it is furthermore identified as an orbital selective Mott transition~\cite{Werner:2009,Ferrero_Cornaglia_Leo_Parcollet_Kotliar_Georges_2009,Gull_Parcollet_Werner_Millis_2009,Kugler_2022} as also suggested by approximations to the Hubbard model~\cite{Worm_Reitner_Held_Toschi_2024}. 

The importance of the Van Hove singularity in the physics of the pseudogap was also identified through DCA calculations~\cite{Wu_Scheurer_Ferrero_Georges_2020} and the dual fermion approximation (DFA) \cite{tanaka2025pseudogap}—a perturbative extension of DMFT designed to incorporate spatial correlations. In the latter work, it was observed that the emergence of pseudogap and strange metal states occurs below a specific hole concentration. This transition point is reached when the renormalized quasiparticle energy at the Van Hove point $\mathbf{k} = (\pi,0)$ is positioned in the immediate vicinity of the Fermi level.

Dynamical Variational Quantum Monte Carlo exhibits the Fermi arcs but does not attempt to correlate them with spin fluctuations~\cite{Rosenberg_Senechal_Tremblay_Charlebois_2022}. 
Inhomogeneous DMFT links the pseudogap to stripe order~\cite{Lichtenstein_Stripes_2001}.

We conclude this section by mentioning that numerical solutions of the Hubbard model are also amenable to a so-called ``fluctuation diagnostic'' of the self-energy.
This approach~\cite{Gunnarsson_Schaefer_LeBlanc_Gull_Merino_Sangiovanni_Rohringer_Toschi_2015,Gunnarsson_Merino_Schaefer_Sangiovanni_Rohringer_Toschi_2018,Schaefer_Toschi_2021,Yu_Iskakov_Gull_Held_Krien_2025} also concludes that spin fluctuations, not superconducting fluctuations, dominate the physics of the pseudogap.
It opens up first at the antinodal point, spreading to the nodal point as the spin correlation length increases.
At large interaction, a resonating valence bond interpretation may become relevant~\cite{Gunnarsson_Merino_Schaefer_Sangiovanni_Rohringer_Toschi_2018}.

The results of essentially all numerical methods mentioned above have been compared in detail for $U=2$ in Ref.~\cite{Schaefer2021}.
The analysis in that paper also concludes that spin fluctuations explain the pseudogap for that value of $U$.
A thorough review of numerical results for the Hubbard model appears in Ref.~\cite{qin2022hubbard}.
%

%
\subsection{Possible relevance to cuprate physics}

Fermi surface topology is clearly relevant to pseudogap physics~\cite{Wu_Scheurer_Chatterjee_Sachdev_Georges_Ferrero_2018,Reymbaut:2019}.
This topology is mainly controlled by the value of the second-neighbor hopping $t'$. 
Nevertheless, for small $t'$, the results of our work should apply. 
In particular, the pseudogap should open up near $(\pi,0)$ on segments of the Fermi surface where $v_F < \gamma/\xi$.
In hole-doped cuprates, the pseudogap is indeed near $(\pi,0)$ and symmetry-related points, so the results of our paper may well be relevant for these compounds.

We are nevertheless cautious because interactions may be too strong for our approach to be valid for hole-doped cuprates.
The physics of strong interactions may be more closely related to superexchange or to gauge theories~\cite{lee2006, Sachdev2020ancillaQbits, Bonetti_Christos_Nikolaenko_Patel_Sachdev_2025, Forni_Bonetti_Müller-Groeling_Vilardi_Metzner_2026}, which are physically different from those considered here.
This is related to the beginning of this discussion section~\ref{sec:Widom}.
Even though this may be just a crossover phenomenon, the physics for weak and for strong interactions may be different, in the same way that Slater and Heisenberg physics are different for antiferromagnetism, even if they are smoothly connected~\cite{senechal2004hot,Hankevych:2006}.

There are reasons to believe from cluster perturbation theory calculations~\cite{senechal2004hot} and also from first-principles calculations~\cite{Weber:2010} that interactions are smaller in electron-doped cuprates. 
It is thus for these compounds that TPSC may be expected to give results directly relevant to experiments.
In electron-doped cuprates, the Fermi surface is far from $(\pi,0)$, so the condition $v_F < \gamma/\xi$ will not be satisfied. 
The pseudogap opens up near the nodal point when the criterion $\xi > \xi_{th}$ is satisfied. 
Near the antinodal points, one recovers a Fermi surface that obeys Fermi-liquid relations~\cite{Gauvin-Ndiaye_Setrakian_Tremblay_2022}, an electron-doped analog of the phenomenon found experimentally in the nodal direction in hole-doped cuprates~\cite{Chang_Mansson_Pailhes_Claesson_Lipscombe_Hayden_Patthey_Tjernberg_Mesot_2013}. 
Theory based on TPSC~\cite{kyung2006pseudogap}, including the above criterion, $\xi > \xi_{th}$, has been verified experimentally~\cite{Armitage_Lu_Kim_Damascelli_Shen_Ronning_Feng_Bogdanov_Shen_Onose_etal._2001,Motoyama_Yu_Vishik_Vajk_Mang_Greven_2007}.
The approach seems to fail near optimal superconducting doping.
This may be due to disorder~\cite{gauvin2022disorder} or, by analogy with what happens in organic conductors~\cite{Bourbonnais_Sedeki_2011}, to interference with superconducting fluctuations.

Even if spin-fermion approaches~\cite{kokkinis2025pseudogap} find agreement with some recent experiments~\cite{Xu_He_Chen_He_Abadi_Rotundu_Lee_Lu_Guo_Tjernberg} in electron-doped cuprates, detailed comparisons with fewer parameters between TPSC and recent experiments on these compounds need to be done~\cite{Horio_Adachi_Mori_Takahashi_Yoshida_Suzuki_Ambolode_Okazaki_Ono_Kumigashira_2016, Horio_Sakai_Suzuki_Nonaka_Hashimoto_Lu_Shen_Ohgi_Konno_Adachi_2025,Xu_He_Chen_He_Abadi_Rotundu_Lee_Lu_Guo_Tjernberg}. 
In particular, we disagree with the statement in Ref.~\cite{Horio_Sakai_Suzuki_Nonaka_Hashimoto_Lu_Shen_Ohgi_Konno_Adachi_2025} that theories based on precursors of antiferromagnetism fail for electron-doped cuprates.  
The argument of that paper is based on a momentum-independent AFM gap.
But in TPSC+ the pseudogap is momentum dependent~\footnote[2]{C. Lahaie and A.-M.S. Tremblay, unpublished.}. 
There are several reviews of ARPES results~\cite{Timusk_Statt_1999,Damascelli:2003,Kordyuk_2015}.

Even though there are qualitative similarities between predicted observables in the weak and strong interaction cases, as we discussed here for the compressibility, several observable quantities can be explained only when interactions are strong, namely, when they overwhelm kinetic energy.

For example, at half-filling, a) the Mott phase, namely an insulating paramagnet with local moments at finite temperature, that is present without any long-range order~\cite{Mott_1949, georges1996, Lee_Nagaosa_Wen_2006}; 
%
%
b) local spin moment $\left< S_z^2\right>$ measured from neutron scattering in agreement with the two-dimensional Heisenberg model~\cite{Keimer_Belk_Birgeneau_1992};

And away from half-filling, for example,
a) X-Ray photoemission spectroscopy that shows a doping-dependent spectrum where the weight of the high-frequency peak decreases with doping roughly half as fast as the weight near the Fermi energy grows~\cite{Fujimori_Takayama-Muromachi_Uchida_Okai_1987, Chen_Sette_Ma_1991, Eskes_Meinders_Sawatzky_1991};
b) resistivity depending linearly on temperature that does not saturate at high temperature, so-called bad-metal behavior, as would be required by the Mott-Ioffe-Regel~\cite{Mott_1972}, namely when the mean-free path becomes comparable to the lattice spacing~\cite{Hussey_2004, Hussey_2008, Deng_Mravlje__2013, Brown_Mitra_Nourafkan_Reymbaut_2018};  
c) kinetic-energy driven antiferromagnetism, or superconductivity as measured by comparing the optical-conductivity sum rule in the normal state with that in the ordered state~\cite{Deutscher_Santander-Syro_Bontemps_2005, Fratino_Semon_Sordi_Tremblay_2016}.

\section{Summary and conclusion}
\label{sec:conclusion}

 In this paper, we applied the TPSC+ approach to the one-band near-neighbor Hubbard model on the square lattice to study the crossover of the pseudogapped electronic liquid to the correlated Fermi liquid as a function of doping. We found, in agreement with recent cold atom experiments~\cite{kendrick2025pseudogap}, that this crossover leads to a maximum in the doping dependence of the compressibility $\kappa(\delta)$ at sufficiently low temperatures, as was first suggested for strong interactions in Ref.~\cite{sordi2012pseudogap}. While the effect is somewhat stronger in relatively small systems, such as those used in cold atom experiments, it persists in the thermodynamic limit as long as the temperature is sufficiently low, $T<0.1$. We predict that at these low temperatures, the maximum in $\kappa(\delta)$ at $\delta_{max}$ exists not only in the intermediate interaction regime $U\approx U_{Mott}$ but also in the weak-to-intermediate interaction regime $U<U_{Mott}$. From the TPSC+ perspective, in both interaction ranges, the maximum is associated with the disappearance of the pseudogap driven by critical thermal spin fluctuations. For these fluctuations, the correlation length diverges as temperature decreases towards an ordered ground state at $T=0$. The pseudogap is thus a precursor to that ordered state.

 We also predict that the uniform magnetic susceptibility $\chi_{sp}(0,0)(\delta)$ exhibits a maximum as a function of doping. At low temperatures $T<0.1$, the doping value at the maximum of $\chi_{sp}(,0,0)(\delta)$ is, practically, the same as the value of $\delta_{max}$ in the compressibility $\kappa(\delta)$. At higher temperatures $0.1<T<0.2$, the uniform magnetic susceptibility $\chi_{sp}(0,0)(\delta)$ still shows a well-defined maximum, whereas the maximum in $\kappa(\delta)$ becomes very broad and eventually disappears. The reason for this difference is that the maximum in $\kappa(\delta)$ requires not only the presence of a pseudogap but also that the width of the Fermi function ($T$) be significantly smaller than the width of the precursor of the lower AFM peak in the density of states. We suggest that in cold atom experiments, the maximum in $\chi_{sp}(,0,0)(\delta)$ may be somewhat easier to identify than the maximum in $\kappa(\delta)=\partial n/\partial\mu$ because it occurs at higher temperatures and does not require numerical differentiation. In addition, we show that TPSC+ correctly predicts a maximum in the temperature dependence of the Knight shift, $\chi_{sp}(0,0)(T)$ at fixed doping, consistent with recent ultracold atom experiments. 

We also established a connection between the appearance of a maximum in $\kappa(\delta)$ and in $\chi_{sp}(0,0)(\delta)$ with the evolution of the single-particle spectra as a function of doping. In particular, the maximum in both quantities occurs when the precursor of the lower SDW peak in the spectral function at the antinodal (Van Hove~\cite{Wu_Scheurer_Ferrero_Georges_2020}) point $\mathbf{k}_{AN}=(\pi,0)$ crosses the line $\omega=0$. Interestingly, the shape of the maximum in both $\kappa(\delta)$ and $\chi_{sp}(0,0)(\delta)$ changes when the critical spin–density-wave fluctuations become strongly incommensurate at very low temperatures: there is an initial rise, followed by a short plateau, then a further rise, and finally a drop. There are two reasons for this behavior: at low temperatures, the pseudogap exists not only at the antinodal point but also at the nodal point, and the multiple peaks in $\chi_{sp}(\mathbf{q},0)$ lead to more than two peaks in both the spectral function and the density of states.

 We compared the compressibility $\kappa=\partial n/\partial \mu$ calculated using the second-level TPSC+ self-energy \eref{eq:selfEnergy2} with $\chi_{ch}(0,0)$ calculated using a momentum and frequency independent $U_{ch}$ at the first level, \eref{eq:chich2} and \eref{eq:sumrule_ch}. In an exact theory, these two quantities must be identical. At a relatively high temperature, $T=0.15$, we found them to be similar. However, at very low temperatures, these quantities differ substantially in the doping region where the compressibility $\kappa(\delta)$ exhibits a maximum. We concluded that in this case, the assumption of a momentum- and frequency-independent effective charge interaction becomes inadequate. By inverting the problem and calculating $U'_{ch}(0,0)$, we found it to be significantly lower than the average $U_{ch}$ obtained from the charge sum rule and Pauli principle. Moreover, in some range of parameters, $U'_{ch}(0,0)$ becomes negative. Negative values of $U'_{ch}(\mathbf{q},0)$ can lead to precursors of SDW-driven charge density waves at finite temperature.

 While our approach is effective within the weak-to-intermediate interaction regime, it is not currently applicable to the strong-coupling limit. In this regime, the physics is dominated by superexchange interactions, which fall outside the scope of our current method.

It is important to emphasize the generality of our result regarding compressibility. Any fluctuating or long-range order that modifies the single-particle spectral weight -- as demonstrated in ~\figref{fig:DSw_vs_delta_U_7_T_0_0714} through \figref{fig:Akw_by_k_U_7_T_0_0250} -- will produce a maximum in the compressibility when entering or leaving that ordered state.
Notably, while this signal manifests in the $\mathbf{q=0}$ charge sector, the underlying fluctuating or long-range order responsible for the compressibility maximum may reside in any channel, including purely spin channels. 
In rhombohedral graphene, for example~\cite{Zhou_Xie_Ghazaryan_Holder_2021}, first-order transitions and Lifshitz transitions are already detected through compressibility measurements. 
Strong antiferromagnetic fluctuations that cannot be detected otherwise in these systems can also give a signal in the compressibility.

Our work opens several promising avenues for future investigation. Notably, it provides a robust framework for a spin-fluctuation interpretation of statistically exact diagrammatic quantum Monte Carlo (DiagMC) results \cite{Simkovic_Rossi_Georges_Ferrero_2024}. Furthermore, our approach enables direct comparisons between future cold-atom experiments and calculations that incorporate realistic parameters such as further-neighbor hopping ($t'$, $t''$) and low temperatures. By facilitating access to parameter regimes relevant to cuprate models—areas where standard quantum Monte Carlo methods can suffer from convergence issues— our work offers a pathway toward a deeper understanding of both the paradigmatic Hubbard model and the physics of high-temperature superconductors.

\section*{Acknowlegments}

 We thank Lev Kendrick, Markus Greiner, Aaron W Young, Alex Deters  and especially Giovanni Sordi for many stimulating discussions and suggestions. We thank Lev Kendrick for sharing compressibility data of the experiments on the cold atom Hubbard model simulator. We thank Thomas Chalopin for sharing QMC, METTS, and cold-atom experimental data regarding the temperature dependence of the Knight shift. 
 We are grateful to Andrey Chubukov and Étienne Lantagne-Hurtubise for discussions. 
 We are indebted to Chloé Gauvin-Ndiaye and Camille Lahaie, who wrote the initial version of the program that we used, and to Antoine Georges and Jérôme Leblanc for useful suggestions.  This work has been supported by the Natural Sciences and Engineering Research Council of Canada (NSERC) under grant RGPIN-2024-05206 and by the Canada First Research Excellence Fund. A.-M.S.T. benefits from RQMP membership https://doi.org/10.69777/309032.

\section*{DATA AVAILABILITY} 

 The data that support the findings of this article are publicly available \cite{VT_data_availability}
 
   
\bibliography{Max_Compress_Biblio}	
\end{document}